\documentclass[aps,showpacs,prb,floatfix,twocolumn,amsmath,citeautoscript]{revtex4-1}
\usepackage{graphicx}
\usepackage{amsmath}
\usepackage{amssymb}
\usepackage{bm}
\usepackage{dcolumn}
\usepackage{subfigure}
\usepackage{units}
\usepackage{hyperref}
\usepackage[usenames]{color}
\usepackage{booktabs}
\usepackage{multirow}
\usepackage{siunitx}
\usepackage{wrapfig}
\hypersetup{pdfborder=0 0 0,colorlinks=true,citecolor=blue,linkcolor=blue}
\input epsf
\newcommand{\rank}{\mathrm{rank}}
\newcommand{\MX}{MX$_2$}
\newcommand{\MS}{MoS$_2$}
\newcommand{\WT}{WTe$_2$}

\newcommand{\BN}{\textit{h}-BN}

\begin{document}
\title{Green's function approach to edge states in transition metal dichalcogenides}
\author{Mojtaba Farmanbar}
\author{Taher Amlaki}
\author{Geert Brocks}
\affiliation{Faculty of Science and Technology and MESA$^+$ Institute for Nanotechnology, University of Twente, P.O. Box 217, 7500 AE Enschede, The Netherlands}
\date{\today}
\begin{abstract}
The semiconducting two-dimensional transition metal dichalcogenides \MX\ show an abundance of one-dimensional metallic edges and grain boundaries.  Standard techniques for calculating edge states typically model nanoribbons, and require the use of supercells. In this paper we formulate a Green's function technique for calculating edge states of (semi-)infinite two-dimensional systems with a single well-defined edge or grain boundary. We express Green's functions in terms of Bloch matrices, constructed from the solutions of a quadratic eigenvalue equation. The technique can be applied to any localized basis representation of the Hamiltonian. Here we use it to
calculate edge states of \MX\ monolayers by means of tight-binding models. Besides the basic zigzag and armchair edges, we study edges with a more general orientation, structurally modifed edges, and grain boundaries. A simple three-band model captures an important part of the edge electronic structures. An eleven-band model comprising all valence orbitals of the M and X atoms, is required to obtain all edge states with energies in the \MX\ band gap. Here states of odd symmetry with respect to a mirror plane through the layer of M atoms have a dangling-bond character, and tend to pin the Fermi level.
\end{abstract}
 
\maketitle
\section{Introduction}
In the wake of graphene a whole new class has emerged of materials that are essentially two-dimensional (2D) in nature.\cite{Geim:natmat07,Xu:chemrev13} The subset of materials with honeycomb-like structures  alone contains metals such as graphene, insulators such as boron nitride (\BN), and semiconductors such as the transition metal dichalcogenides (\MX; M = Mo,W, X = S,Se,Te).\cite{Geim:nat13,Chhowalla:natchem13} It becomes more and more feasible to grow nanostructures and in-plane heterostructures of 2D materials in a controlled way.\cite{Helveg:prl00,Ci:natmat:10,Levendorf:nat12,vanderZande:natmat13,Najmaei:natmat13,Liu:prl14,Li:science15} As a result the electronic structure of edges and grain boundaries attracts increasing attention. 

Graphene edges, for instance, are predicted to have remarkable one-dimensional electronic and magnetic properties.\cite{Son:nat06,Yazyev:prl08} The edges and grain boundaries of \MX\ sheets are generally metallic.\cite{Bollinger:prl1,Bollinger:prb3,Vojvodic:prb9,Botello:nanotech9,Andersen:prb14,Zou:nanol12,Zhou:nanol14,Liu:prl14} As the bulk 2D \MX\ materials are semiconducting, the metallicity is truly localized at the edge or grain boundary, where one could see manifestations of the peculiar spectral and transport properties of one-dimensional (1D) metals \cite{Giamarchi:03}. In chemistry \MS\ edges have been identified as sites that show a special catalytic activity.\cite{Bollinger:prl1,Vojvodic:prb9,Schweiger:jcat02} Such experimental developments have motivated a large number of calculations on the electronic structure of edge states, both first-principles calculations, as well as model calculations.   

The electronic structure of graphene edges in particular have been studied extensively. As graphene has a relatively simple electronic structure, some features of the edge states in graphene can be studied by analytical or simple numerical techniques.\cite{Nakada:prb96,Akhmerov:prb08,Delplace:prb11} The edges of \MS\ have attracted renewed attention recently.\cite{Bollinger:prl1,Bollinger:prb3,Vojvodic:prb9,Botello:nanotech9,Andersen:prb14,Ataca:jpcc10,Wang:jacs2010,Zou:nanol12,Kou:jpcl12,Fernandez:jmatres12,Cakir:prb89,Chu:prb14,Zhou:nanol14,Xu:prb14,Lucking:cm15,Pavlovic:prb15} Here the complexity of the electronic structure requires more extensive calculations, even for relatively simple modeling at the tight-binding level. A standard technique for calculating edge states uses supercells to model nanoribbons of a finite width. Drawbacks of this approach are that the electronic structure of the edge states is mixed with that of the bulk-like interior of the nanoribbon, and that the two edges of the nanoribbon can interact electronically. The ribbon has therefore to be sufficiently wide in order to electronically separate the two edges from one another and from the bulk. It may require the use of large supercells, which in particular if the materials are modeled from first principles, leads to time-consuming calculations and complicates analysis of the results.

Green's function techniques constitute an alternative approach. They enable calculations on semi-infinite structures with a single well-defined edge, or on infinite structures containing a single grain boundary, and they generally do not require the use of large supercells. Green's function techniques have been pioneered for calculations on surface states of 3D materials.\cite{Bernholc:prb78,Pollmann:prb78,Lee:prb81a,Lee:prb81b,Chang:prb82} Here we formulate a special Green's function technique for calculating edge and grain boundary states. This technique is inspired by the Green's function formalism that has been introduced for calculating electronic transport through nanostructures.\cite{Ando:prb91,Khomyakov:prb04,Khomyakov:prb05,Khomyakov:prb06,Zwierzycki:pssb07} 

We express both edge Green's functions and bulk Green's functions in terms of Bloch matrices, which are constructed from the eigenvalues and eigenfunctions of a quadratic eigenvalue problem.\cite{Ando:prb91,Tisseur:SIAM01,Khomyakov:prb04,Khomyakov:prb05,Khomyakov:prb06,Zwierzycki:pssb07} Structural and chemical modifications at the edges and grain boundaries are then tackled by connecting the modified edges to semi-infinite bulk structures. The method allows for a clean separation of edge and bulk properties at a moderate computational cost.

Our Green's function approach requires a representation of the Hamiltonian on a localized basis, such as atomic orbitals or Wannier functions, or a real space grid. It can be applied to tight-binding, as well as first-principles models. In this paper we illustrate its use on tight-binding models for \MX\ monolayers, \MS\ in particular. We apply the approach to the three-band model for the simplified electronic structure of \MS, developed by Mattheis\cite{Mattheis:prb73} and Liu \textit{et al.}.\cite{Liu:prb13} We study the electronic structures of the elementary (zigzag, armchair) \MS\ edges, and of edges of a more general orientation. We illustrate the effect on the electronic structure of edge modifications and of the formation of grain boundaries. For comparison we also study edges within an eleven-band model comprising all \MS\ valence orbitals, developed by Cappelluti \textit{et al.}\cite{Cappel:prb88}.   

The paper is organized as follows. In Sec.~\ref{sec:green} we formulate the technique of calculating Green's functions using Bloch matrices. How to apply this technique to \MX\ edges and grain boundaries is described in Sec.~\ref{sec:edges}. The tight-binding models are discussed in Sec.~\ref{sec:tbmodels}, and appendices~\ref{sec:3band} and \ref{sec:11band}. We discuss results obtained with these models for \MX\ edges and grain boundaries in Secs.~\ref{sec:result1} and \ref{sec:result2}.

\section{Theory}

\subsection{Green's functions}
\label{sec:green}
\begin{figure}[t]
\includegraphics[width=5.5cm]{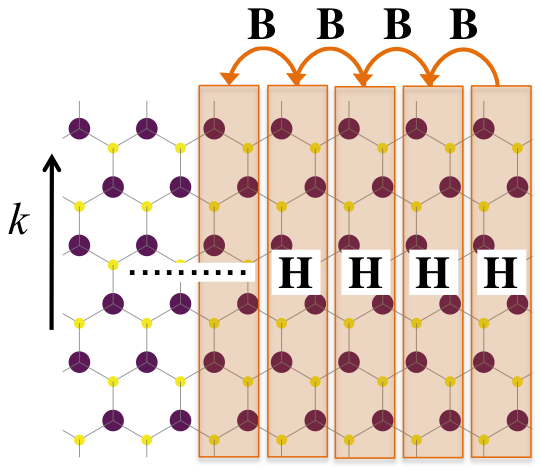}
\caption{(Color online) A 2D layer is divided in 1D strips. Translation symmetry along the strips gives the Bloch wave number $k$. Translation symmetry between the strips results in identical on-strip Hamiltonian matrix blocks $\mathbf{H}(k)$ and hopping matrix blocks $\mathbf{B}(k)$.  
}
\label{fig:slices}
\end{figure}

We divide the 2D layers into 1D strips, see Fig.~\ref{fig:slices}.\cite{Lee:prb81a} Assuming translational symmetry with period $a$ along the strip, the Hamiltonian matrix can be labeled by a Bloch wave number $-\frac{\pi}{a} <k \leq \frac{\pi}{a}$. For clarity of notation we suppress the label $k$ in the following. The thickness of the strips is chosen such that a direct interaction exists between neighboring strips only. We label the strips by an index $i$, and divide the Hamiltonian matrix into blocks $\mathbf{H}_{i,j}$, with an $i,j=-\infty,\dots,\infty$ for an inifinite system. Having only nearest neighbor interactions between strips means that the Hamiltonian matrix is block tridiagonal, i.e., $\mathbf{H}_{i,j}=0;j\neq i,i\pm 1$. For a unit cell in the strip containing $N$ orbitals, all these matrix blocks are $N\times N$. 

The columns of the retarded Green's function matrix blocks $\mathbf{G}_{i,j}$ obey 
\begin{equation}
\label{eq:g1}
-\mathbf{H}_{i,i-1}\mathbf{G}_{i-1,j}+\left(E^+\mathbf{I}-\mathbf{H}_{i,i}\right)\mathbf{G}_{i,j}-\mathbf{H}_{i,i+1} \mathbf{G}_{i+1,j}=\mathbf{I}\delta_{ij},
\end{equation}
where $\mathbf{I}$ is the $N\times N$ identity matrix, and $E^+=E+i\eta$ with $\eta$ the usual infinitesimal positive real number. Note that we assume a representation based upon an orthogonal basis set. We are foremost interested in the layer resolved density of states, given by the usual expression $n_i=-\pi^{-1}\mathrm{ImTr}\mathbf{G}_{i,i}$.
Obviously, besides on the Bloch wave number $k$, the Green's function matrix also depends on the energy $E$. Again for ease of notation we often omit both these labels in the following.

In an infinite system with translational symmetry between the layers, the strips are identical, and Eq.~\ref{eq:g1} becomes
\begin{equation}
\label{eq:g2}
-\mathbf{B}\mathbf{G}_{i-1,j}+\left(E^+\mathbf{I}-\mathbf{H}\right)\mathbf{G}_{i,j}-\mathbf{B}^\dagger \mathbf{G}_{i+1,j}=\mathbf{I}\delta_{ij},
\end{equation}
with $\mathbf{B}=\mathbf{H}_{i,i-1}$, $\mathbf{H}=\mathbf{H}_{i,i}$, and $\mathbf{B}^\dagger=\mathbf{H}_{i,i+1}$ for $i=-\infty,\dots,\infty$, see Fig.~\ref{fig:slices}. For a left semi-infinite system with $i=-\infty$ to $i=0$, the equations remain the same, except for the $i=0$ equation, which becomes  
\begin{equation}
\label{eq:g3}
-\mathbf{B}\mathbf{G}^{L}_{-1,j}+\left(E^+\mathbf{I}-\mathbf{H}\right)\mathbf{G}^{L}_{0,j}=\mathbf{I}\delta_{0j}.
\end{equation}
In surface science the matrix block $\mathbf{g}_L=\mathbf{G}^{L}_{0,0}$ is called the surface Green's function. In the context of 2D structures we call it the edge Green's function. 

In a similar way, for a right semi-infinite system with $i=0,\infty$, only the equation for $i=0$ is different from the bulk equation, Eq.~\ref{eq:g2}
\begin{equation}
\label{eq:g4}
\left(E^+\mathbf{I}-\mathbf{H}\right)\mathbf{G}^{R}_{0,j}-\mathbf{B}^\dagger \mathbf{G}^{R}_{1,j}=\mathbf{I}\delta_{0j},
\end{equation}
with the matrix block $\mathbf{g}_R=\mathbf{G}^{R}_{0,0}$ the edge Green's function. Note that if the 2D layer has no two-fold symmetry, such as inversion, a mirror plane perpendicular to the 2D layer, or a two-fold rotation axis perpendicular to the 2D layer, then $\mathbf{g}_R \neq \mathbf{g}_L$, i.e., a right edge is different from a left edge.

The density of states of the edge strip of a left/right semi-infinite system is given by
\begin{equation}
\label{eq:g3a}
n_{L/R}^{(S)}=\sum_{\alpha=1}^N n_\alpha;\;\;\; n_\alpha=-\pi^{-1}\mathrm{Im}\left[ \mathbf{g}_{L/R} \right]_{\alpha,\alpha},
\end{equation}
where $n_\alpha$ is the density of states projected on a single orbital $\alpha$.

Partitioning the infinite system into a right and a left semi-infinite halve, the on-strip matrix blocks of the Green's function of the infinite system can be expressed as
\begin{equation}
\label{eq:g5}
\mathbf{G}_{i,i}=\left[\mathbf{g}_L^{-1} - \mathbf{B}^\dagger\mathbf{g}_R\mathbf{B} \right]^{-1}.
\end{equation}

\subsubsection{Eigenmodes and Bloch matrices}
We will express the Green's function matrices of (semi)infinite systems in terms of Bloch matrices.\cite{Ando:prb91} Equation~\ref{eq:g2} for $i\neq j$ is the same as the Schr{\"{o}}dinger equation in tight-binding representation
\begin{equation}
\label{eq:e1}
-\mathbf{B}\mathbf{c}_{i-1}+\left(E\mathbf{I}-\mathbf{H}\right)\mathbf{c}_{i}-\mathbf{B}^{\dagger}\mathbf{c}_{i+1}=0,
\end{equation}
where $\mathbf{c}_{i}$ is the $N$-dimensional vector of orbital coefficients of the wave function on strip $i$. With translational symmetry between the strips, the elementary solution is a Bloch wave, and $\mathbf{c}_{i+1}=\lambda\mathbf{c}_{i}$, with $\lambda$ a complex constant. 
For an inifinite system this holds for all $i$, and for a left (right) semi-infinite system for $i<0$ ($i>0$). Using this relation as an ansatz in Eq.~\ref{eq:e1} gives a quadratic eigenvalue equation in $\lambda$ of dimension $N$. 
\begin{equation}
\mathbf{A}(\lambda)\mathbf{u}=0;\;\; \mathbf{A}(\lambda) = -\mathbf{B}+\lambda \left(E\mathbf{I}-\mathbf{H}\right)-\lambda^2\mathbf{B}^{\dagger}.
\label{eq:e3a}
\end{equation}
As $\mathbf{A}(\lambda) = \lambda^2\mathbf{A}^\dagger(1/\lambda^*)$, if $\lambda$ is a root of $\det\mathbf{A}(\lambda)$, then so is $1/\lambda^*$. Numerically, the solutions are usually found by solving an equivalent linear generalized eigenvalue equation of dimension $2N$,\cite{Tisseur:SIAM01}
\begin{equation}
\label{eq:e3}
\left[\left(\begin{array}{cc}
\mathbf{I} & \mathbf{0} \\
E\mathbf{I}-\mathbf{H} & -\mathbf{B}
\end{array}\right)-\lambda\left(\begin{array}{cc}
\mathbf{0} & \mathbf{I} \\
\mathbf{B}^{\dagger} & \mathbf{0}
\end{array}\right)\right]\mathbf{z}=
\mathbf{0},
\end{equation}
which resembles the eigenvalue equation for the transfer matrix.\cite{Lee:prb81a,Lee:prb81b,Hatsugai:prl93,Hatsugai:prb08,Teo:prb08,Mong:prb11} 

The maximally $2N$ solutions of this equation can be divided into two classes, i.e. right-going modes and left-going modes, labeled respectively by $+$ and $-$ superscripts in the following. Right-going modes are either evanescent waves that decay to the right, $|\lambda_{n}^{+}|<1$, or waves that travel to the right, meaning $|\lambda_{n}^{+}|=1$ and a positive group velocity. Left-going modes either decay to the left, $|\lambda_{n}^{-}|>1$, or travel to the left, $|\lambda_{n}^{-}|=1$ with negative group velocity. With $\mathbf{u}_n$ the eigenvector belonging to the eigenvalue $\lambda_n$ in Eq.~\ref{eq:e3}, the group velocity is given by\cite{Khomyakov:prb05}
\begin{equation}
\label{eq:e4}
v_n = -\frac{2a}{\hbar}\mathrm{Im}\left[\lambda_n\mathbf{u}_n^\dagger \mathbf{B}^\dagger \mathbf{u}_n \right],
\end{equation}
Only for traveling waves is the group velocity non-zero. 

We divide the eigenvectors into a set of $N^+$ right-going modes $\mathbf{u}^+_n$ and a set of $N^-$ left-going modes $\mathbf{u}^-_n$. The evanescent waves always come in pairs of a right-going and a left-going mode, i.e., if $\lambda^+_n$ is the eigenvalue of a right-going mode, then $\lambda^-_n=1/(\lambda^+_n)^*$ gives a left-going solution. Traveling waves do not necessarily come in such pairs, and the numbers of right- and left-going traveling waves may be different. Neither right-going modes or left-going modes necessarily form a complete set in $N$-dimensional space, nor are they an orthogonal set in general.

One can use these two sets of modes to form the two $N\times N^\pm$ matrices 
\begin{equation}
\label{eq:e5}
\mathbf{U}^\pm=\left(\mathbf{u}_1^\pm,\mathbf{u}_1^\pm,\dots,\mathbf{u}_{N^\pm}^\pm \right),
\end{equation}
and construct the two Bloch matrices
\begin{equation}
\label{eq:e6}
\mathbf{F}^\pm=\mathbf{U}^\pm\mathbf{\Lambda}^\pm\widetilde{\mathbf{U}}^\pm,
\end{equation}
where $\mathbf{\Lambda}^\pm$ are the $N^\pm \times N^\pm$ diagonal matrices with the eigenvalues $\lambda_n^\pm$ on the diagonal, and $\widetilde{\mathbf{U}}^\pm$ are the $N^\pm \times N$ (pseudo)inverses of $\mathbf{U}^\pm$.\cite{Golub:96} The Bloch matrices have the convenient property 
\begin{equation}
\label{eq:e7}
\left(\mathbf{F}^\pm\right)^p=\mathbf{U}^\pm\left(\mathbf{\Lambda}^\pm\right)^p\widetilde{\mathbf{U}}^\pm,
\end{equation}
for any integer $p$, as $\widetilde{\mathbf{U}}^\pm\mathbf{U}^\pm=\mathbf{I}_{M^\pm}$, the $M^\pm\times M^\pm$ identity matrix, with $M^\pm\leq N^\pm$. It follows that
\begin{equation}
\label{eq:e8}
\mathbf{c}_i = \left(\mathbf{F}^+\right)^i \mathbf{c}_0^+ + \left(\mathbf{F}^-\right)^i \mathbf{c}_0^-,
\end{equation}
satisfies the tight-binding equation, Eq.~\ref{eq:e1}, where $\mathbf{c}_0^\pm$ set the boundary conditions in strip number 0. We assume that all relevant solutions can be expressed this way.  
 
\subsubsection{Green's functions in terms of Bloch matrices}
The general expression of Eq.~\ref{eq:e8} also applies to the columns of the Green's function matrices, compare Eqs.~\ref{eq:g2} and \ref{eq:e1}. The boundary conditions require that a retarded Green's function comprises traveling waves moving outwards from its point source and/or evanescent waves decaying away from the source. For left and right semi-infinite systems this gives
\begin{eqnarray}
\label{eq:b1}
\mathbf{G}_{i,0}^{L} &=& \left(\mathbf{F}^-\right)^i \mathbf{g}_L, \; i<0, \\ 
\label{eq:b1a}
\mathbf{G}_{i,0}^{R} &=& \left(\mathbf{F}^+\right)^i \mathbf{g}_R, \; i>0.
\end{eqnarray} 
Using Eq.~\ref{eq:b1} in Eqs.~\ref{eq:g2} and \ref{eq:g3} then leads to
\begin{eqnarray}
\label{eq:b2}
&&\left[ -\mathbf{B}\left(\mathbf{F}^-\right)^{-1} + E^+\mathbf{I} - \mathbf{H} \right]\mathbf{g}_L = \mathbf{I}, \\ \nonumber
&&\left[ -\mathbf{B}\left(\mathbf{F}^-\right)^{-1} + E^+\mathbf{I} - \mathbf{H} -\mathbf{B}^\dagger\mathbf{F}^- \right]\left(\mathbf{F}^-\right)^{-1}\mathbf{g}_L = \mathbf{0}. 
\end{eqnarray} 
Solving these two equations gives the edge Green's function of a left semi-infinite system as
\begin{equation}
\label{eq:b3}
\mathbf{g}_L = \left[\mathbf{B}^\dagger\mathbf{F}^-\right]^{-1},
\end{equation}  
where the inversion should be treated as a pseudoinversion if $\mathbf{B}^\dagger$ is singular.\cite{Golub:96} Using Eq.~\ref{eq:b1a} in Eqs.~\ref{eq:g2} and \ref{eq:g4} gives the edge Green's function of a right semi-infinite system 
\begin{equation}
\label{eq:b4}
\mathbf{g}_R = \left[\mathbf{B}\left(\mathbf{F}^+\right)^{-1}\right]^{-1}.
\end{equation}  
Finally, using Eq.~\ref{eq:g5} gives the on-strip Green's function matrix block of an infinite system
\begin{equation}
\label{eq:b5}
\mathbf{G}_{i,i}=\left[\mathbf{B}^\dagger\mathbf{F}^- - \mathbf{B}^\dagger\mathbf{F}^+ \right]^{-1}.
\end{equation}

\subsubsection{Ideal edge states} 
One cannot have traveling Bloch waves for energies in the band gap of a semiconductor. In semi-infite systems one can however have solutions in the form of evanescent states that originate from the edge of the system. One can find the energies of these edge states from the edge Green's functions, Eqs.~\ref{eq:b3} and \ref{eq:b4}, which have isolated poles at these energies. Obviously in numerical calculations one has to work at complex energies $E+i\eta$ to avoid these poles, but $\eta$ can be chosen small.

Alternatively, edge states can be obtained from the solutions of the eigenvalue problem, Eq.~\ref{eq:e3a}, solved at real energies $E$.\cite{Lee:prb81a,Harrison:ps03,Teo:prb08,Dang:prb14} As edge states should decay away from the edge, only the $\mathbf{u}_n^-$ modes can contribute to an edge state for a left semi-infinite system, and only the $\mathbf{u}_n^+$ modes for a right semi-infinite system. An edge state of a left semi-infinite system has amplitude zero beyond the edge of that system, i.e., $\mathbf{c}_1=\mathbf{F}^-\mathbf{c}_0^-=0$ (Eq.~\ref{eq:e8}). This means $\rank(\mathbf{F}^-)<N^-$, the number of left-going solutions of Eq.~\ref{eq:e3a}. Because $\rank(\mathbf{F}^-)=\rank(\mathbf{U}^-)$, see Eq.~\ref{eq:e6} with $\rank(\Lambda^-)=N^-$ and  $\rank(\mathbf{U}^-)=\rank(\widetilde{\mathbf{U}}^-)$, a necessary and sufficient condition for the existence of an edge state is that the eigenmodes $\mathbf{u}_n^-$ are linearly dependent. A similar reasoning holds for the edge states of a right semi-infinite system. The number of edge states at a particular energy $E$ and wave number $k$ of a left/right semi-infinite system is then given by $N_\mathrm{edge}^{L/R}=N^{-/+}-\rank(\mathbf{U}^{-/+})$. 

\subsubsection{Ideal grain boundaries}

\begin{figure}[t]
\includegraphics[width=6.0cm]{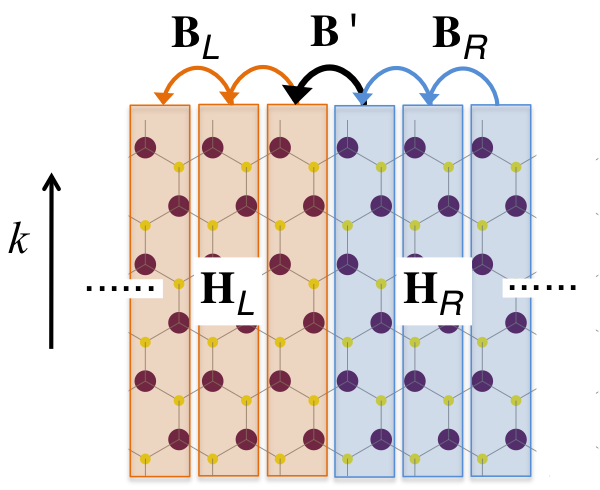}
\caption{(Color online) Grain boundary between a left semi-infinite system with on-strip Hamiltonian matrix blocks $\mathbf{H}_L(k)$ and hopping matrix blocks $\mathbf{B}_L(k)$ and a right semi-infinite system with matrix blocks $\mathbf{H}_R(k)$ and $\mathbf{B}_R(k)$. The coupling between left and right parts is given by the hopping matrix block $\mathbf{B}'(k)$.  
}
\label{fig:grainboundary}
\end{figure}

A model for an ideal grain boundary is shown in Fig.~\ref{fig:grainboundary}. Space is divided into two parts with $\mathbf{B}'$ the hopping matrix block connecting the left and right halves. We assume that the on-strip Hamiltonian matrix blocks of all the strips in the left half are given by $\mathbf{H}_L$ right up to the boundary, and that the hopping matrix blocks between all nearest neighbor strips in the left part are given by $\mathbf{B}_L$. The corresponding matrix blocks for the right half are $\mathbf{H}_R$ and $\mathbf{B}_R$, respectively. The Green's function matrix blocks $\mathbf{g}_L^I$ and $\mathbf{g}_R^I$ pertaining to the two strips just left and right of the grain boundary interface can be derived like Eq.~\ref{eq:g5} 
\begin{eqnarray}
\label{eq:s3}
\mathbf{g}_L^I&=&\left[\mathbf{g}_L^{-1} - \mathbf{B'}^\dagger\mathbf{g}_R\mathbf{B'} \right]^{-1}, \\
\label{eq:s3a}
\mathbf{g}_R^I&=&\left[\mathbf{g}_R^{-1} - \mathbf{B'}\mathbf{g}_L\mathbf{B'}^\dagger \right]^{-1},
\end{eqnarray}
where $\mathbf{g}_L$ and $\mathbf{g}_R$ are the edge Green functions of the left and right semi-infinite systems, respectively. With Eqs.~\ref{eq:b3} and \ref{eq:b4} one can express the interface Green's functions in terms of Bloch matrices
\begin{eqnarray}
\label{eq:s4}
\mathbf{g}_L^I&=&\left[\mathbf{B}_L^\dagger\mathbf{F}^-_L - \mathbf{B'}(\mathbf{B}_R(\mathbf{F}^+_R)^{-1})^{-1}\mathbf{B'}^\dagger \right]^{-1}, \\
\mathbf{g}_R^I&=&\left[\mathbf{B}_R(\mathbf{F}^+_R)^{-1} - \mathbf{B'}(\mathbf{B}_L\mathbf{F}^-_L)^{-1}\mathbf{B'}^\dagger \right]^{-1}.
\end{eqnarray}

\subsubsection{Modified edges}

\begin{figure}[t]
\includegraphics[width=5.5cm]{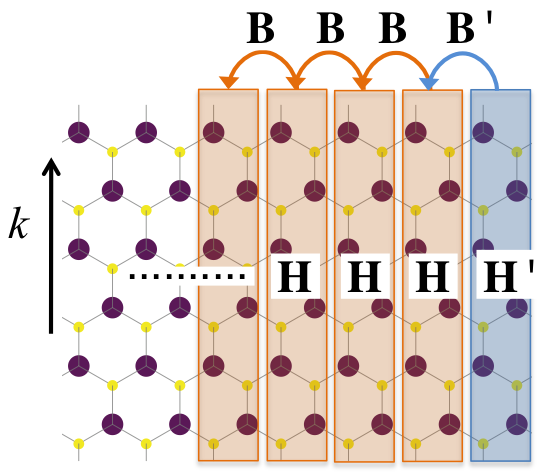}
\caption{(Color online) Left semi-infinite system with a modified edge. $\mathbf{H}(k)$ and $\mathbf{B}(k)$ are the matrix blocks of the unperturbed system; $\mathbf{H}'(k)$ and $\mathbf{B}'(k)$ are the matrix blocks of the perturbed edge strip.  
}
\label{fig:modedge}
\end{figure}

\label{sec:modedge}
So far we have assumed that all layers are identical right up to an edge or grain boundary. The formation of an edge or boundary often involves an electronic and a structural reconstruction, which makes the Hamiltonian matrix blocks of the strips adjacent to an edge or grain boundary different from those of the bulk strips. 

We illustrate this on a left semi-infinite system with an edge layer $(i=0)$ that is different from the bulk. The tight-binding equations for the two layers closest to the edge are
\begin{eqnarray}
\label{eq:s5}
-\mathbf{B}\mathbf{c}_{-2}+\left(E\mathbf{I}-\mathbf{H}\right)\mathbf{c}_{-1}-
\mathbf{B'}^{\dagger}\mathbf{c}_{0}&=&0 \nonumber \\
-\mathbf{B'}\mathbf{c}_{-1}+\left(E\mathbf{I}-\mathbf{H'}\right)\mathbf{c}_{0}&=&0,
\end{eqnarray}
where $\mathbf{H'}\neq \mathbf{H}$ is the on-site Hamiltonian of the edge layer, and $\mathbf{B'}\neq \mathbf{B}$ is the coupling of this layer to the rest of the system, see Fig.~\ref{fig:modedge}. Using $\mathbf{c}_{-2}=(\mathbf{F}^{-})^{-1}\mathbf{c}_{-1}$, see Eqs.~\ref{eq:e8} and \ref{eq:b1}, transforms Eq.~\ref{eq:s5} into
\begin{equation}
\label{eq:s6}
\left(\begin{array}{cc}
E\mathbf{I}-\mathbf{H}-\mathbf{B}(\mathbf{F}^{-})^{-1} & -\mathbf{B'}^{\dagger}\\
-\mathbf{B'} & E\mathbf{I}-\mathbf{H'}
\end{array}\right)\left(\begin{array}{c}
\mathbf{c}_{-1}\\
\mathbf{c}_{0}
\end{array}\right)=\left(\begin{array}{c}
\mathbf{0}\\
\mathbf{0}
\end{array}\right).
\end{equation}
In the terminology used in non-equilibrium Green's function (NEGF) transport calculations, the term $\boldsymbol{\Sigma}_{L}(E)\equiv \mathbf{B}(\mathbf{F}^{-})^{-1}$ is called the self-energy of the left lead.\cite{Khomyakov:prb05} The Green's function matrix blocks pertaining to the two layers closest to the edge are then given by
\begin{equation}
\label{eq:s7}
\mathbf{G}_L=\left(\begin{array}{cc}
E\mathbf{I}-\mathbf{H}-\mathbf{B}(\mathbf{F}^{-})^{-1} & -\mathbf{B'}^{\dagger}\\
-\mathbf{B'} & E\mathbf{I}-\mathbf{H'}
\end{array}\right)^{-1}.
\end{equation}

The Green's function gives the density of states in the usual way, cf. Eq.~\ref{eq:g3a}. The density of states is zero for energies inside the band gap, except for isolated poles at particular energies, which represent the edge states of the modified edges. This formalism can be adapted in an obvious way to model modified edges of right semi-infinite systems, or grain boundaries where the strips left and right of the interface are modified.

\subsubsection{Charge neutrality level}
Experimentally the Fermi level in \MX\ compounds is often determined by unintentional doping due to defects.\cite{Diana:prb08,McDonnell:acs8,Santosh:Nanotech14} In addition, as \MX\ compounds are polar materials, an internal electric field is created if a crystallite is terminated by polar edges.\cite{Guller:prb13,Gilbertini:nl15} Such an electric field can cause a long range charge transfer between different edges, even if the bulk material does not contain any impurities. The position of the Fermi level can then become dependent on the size and the shape of the sample, which makes the intrinsic Fermi level an ill-defined quantity. 

Each edge or grain boundary has a well-defined energy level at which that edge or grain boundary is electrically neutral, the charge neutrality level (CNL). We use the CNL as a reference point in the following. The charge neutrality level $E_\mathrm{CNL}$ is defined as the energy at which
\begin{equation}
\label{eq:s8}
 N(E_\mathrm{CNL})=N_\mathrm{strip},
\end{equation}
where $N_\mathrm{strip}$ is the number of electrons that makes the strip neutral, and $N(E)$ is the electron counting function
\begin{equation}
\label{eq:s9}
N(E)=\int_{-\infty}^{E}n(E')dE',
\end{equation}
with $n(E)$ is the $k$-integrated density of states
\begin{equation}
\label{eq:s10}
n(E)=\frac{a}{2\pi}\int_{-\frac{\pi}{a}}^{\frac{\pi}{a}}n(E,k)dk.
\end{equation}
The $k$-resolved density of states $n(E,k)$ can be obtained from Eq.~\ref{eq:g3a} for edge strips, and a similar expression for a bulk strip.

%
\subsection{\MX\ edges}
\label{sec:edges}

The hexagonal lattice of \MX\ is shown in Fig.~\ref{fig:lat}. We specify an edge starting from a supercell spanned by vectors $\mathbf{T}_{1}$ and $\mathbf{T}_{2}$. This supercell is used to define a semi-infinite system, choosing one of the vectors as the translation vector parallel to the edge.

\subsubsection{Zigzag and armchair edges}
\begin{figure}[t]
\includegraphics[width=6.0cm]{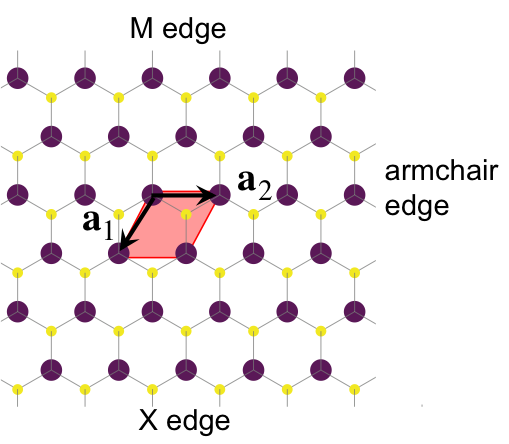}
\caption{(Color online) Top view of the \MX\ lattice with M and X atoms in purple and yellow, respectively. The lattice vectors $\mathbf{a}_1$ and $\mathbf{a}_2$ are indicated by arrows. The M and X zigzag edges runs parallel to $\mathbf{a}_2$ and are the edges of right and left semi-infinite systems, respectively. The armchair edge runs parallel to $2\mathbf{a}_1+\mathbf{a}_2$. 
}
\label{fig:lat}
\end{figure}

Similar to graphene the basic-type edges of the \MX\ lattice are the zigzag and armchair edges as defined in Fig.~\ref{fig:lat}. A zigzag edge is defined by $\mathbf{T}_{1}=\mathbf{a}_{1}$ and $\mathbf{T}_{2}=\mathbf{a}_{2}$, with $\mathbf{T}_{2}$ as the vector parallel to the edge. The matrix blocks discussed in section~\ref{sec:green} become
\begin{eqnarray}
\label{eq:z1}
\mathbf{H} &=& \mathbf{H}_{0,0}+\mathbf{H}_{0,1}e^{i2\pi k}+\mathbf{H}_{0,-1}e^{-i2\pi k}, \\
\label{eq:z1a}
\mathbf{B} &=& \mathbf{H}_{-1,0}+\mathbf{H}_{-1,-1}e^{-i2\pi k},
\end{eqnarray}
where $\mathbf{H}_{p,q}$ denotes the real space Hamiltonian matrix block that describes the interaction between atoms in the unit cell situated at the origin and atoms in the unit cell situated at $p\mathbf{a}_1+q\mathbf{a}_2$. The matrix elements of $\mathbf{H}_{p,q}$ depend on the specific tight-binding model that is used to represent the electronic structure of \MX. They are given in the appendices.

Note that by solving the quadratic eigenvalue equation, Eqs.~\ref{eq:e3a} and \ref{eq:e3}, one has simultaneous access to the Green's functions of the edges of a right and a left semi-infinite system via Eqs.~\ref{eq:b3} and \ref{eq:b4}. Unlike graphene the \MX\ monolayer lacks inversion symmetry, which means that the zigzag edge termination of a right semi-infinite system is different from that of a left semi-infinite system, see Fig.~\ref{fig:lat}. The zigzag edge of a right semi-infinite system is terminated by metal atoms. We call this the M-edge, consistent with previous studies on \MS, where it is called the Mo-edge.\cite{Bollinger:prl1,Bollinger:prb3,Vojvodic:prb9} The zigzag edge of a left semi-infinite system is then called the X-edge, as it is terminated by chalcogen atoms. In previous studies on \MS, it has been called the S-edge. 

An edge in armchair orientationcan be constructed from $\mathbf{T}_{1}=2\mathbf{a}_{1}+\mathbf{a}_{2}$ and $\mathbf{T}_{2}=\mathbf{a}_{2}$ with $\mathbf{T}_{1}$ the translation vector parallel to the edge, see Fig.~\ref{fig:lat}. The corresponding supercell is twice as large as the unit cell, so the dimension of the matrix blocks defining the edge are twice the dimension of the blocks defining the zigzag edge.
\begin{equation}
\label{eq:z3}
\mathbf{H}=\left(\begin{array}{cc}
\mathbf{H}_{0,0} & \mathbf{H}_{1,1}+\mathbf{H}_{-1,0}e^{-i2\pi k}\\
\mathbf{H}_{1,0}e^{i2\pi k}+\mathbf{H}_{-1,-1} & \mathbf{H}_{0,0}
\end{array}\right),
\end{equation}
\begin{equation}
\label{eq:z4}
\mathbf{B}=\left(\begin{array}{cc}
\mathbf{H}_{0,-1} & \mathbf{H}_{1,0}+\mathbf{H}_{-1,-1}e^{-i2\pi k}\\
0 & \mathbf{H}_{0,-1}
\end{array}\right).
\end{equation}
Note that the termination at the edge is controlled by the contents of the cell used to define the primitive vectors $\mathbf{a}_{1},\mathbf{a}_{2}$. In particular, the cell defined in Fig.~\ref{fig:lat} does not lead to a pristine armchair edge, but one with additional M or X atoms attached to the edge. It is straightforward to remove these atoms and by applying the technique outlined in Sec.~\ref{sec:modedge} obtain the electronic structure of a pristine armchair edge. 

\subsubsection{General edges}
\label{sec:genedge}

\begin{figure}[t]
\includegraphics[width=9.0cm]{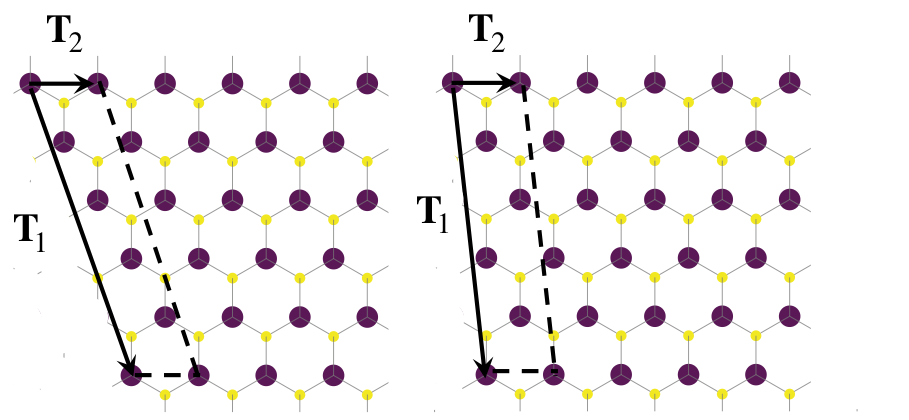}
\caption{(Color online) Left: generalized zigzag edge parallel to $\mathbf{T}_1=3(\mathbf{a}_{1}+\mathbf{a}_{2})+(2\mathbf{a}_{1}+\mathbf{a}_{2})$. Right: generalized armchair edge parallel to $\mathbf{T}_1=(\mathbf{a}_{1}+\mathbf{a}_{2})+2(2\mathbf{a}_{1}+\mathbf{a}_{2})$.}
\label{fig:super}
\end{figure}

Edges with a somewhat more general orientation are defined by the supercell
\begin{equation}
\label{eq:z6a}
\mathbf{T}_{1}=m\left(\mathbf{a}_{1}+\mathbf{a}_{2}\right)+n\left(2\mathbf{a}_{1}+\mathbf{a}_{2}\right);\; \mathbf{T}_{2}=\mathbf{a}_{2}, 
\end{equation}
with $\mathbf{T}_{1}$ the translation vector parallel to the edge, defined as $m$ zigzag vectors plus $n$ armchair vectors, see Fig.~\ref{fig:super}. The angle with the direction of $\mathbf{a}_{2}$ is given by
\begin{equation}
\label{eq:z6}
\theta=\arccos\left(\frac{\frac{1}{2}m}{\sqrt{m^{2}+3mn+3n^{2}}}\right).
\end{equation}
Because of the symmetry of the lattice one only has to cover the $30^{\mathbf{o}}$ angle between a zigzag orientation $m=1,n=0$, $\theta=60^{\mathrm{o}}$, and an armchair orientation $m=0,n=1$, $\theta=90^{\mathrm{o}}$. Left and right edges (M-type and X-type edges) are then obtained by using $\mathbf{T}_2=\mathbf{a}_2$ and $\mathbf{T}_2=-\mathbf{a}_2$, respectively. 

A series of edge orientations is obtained by setting $n=1$ and varying $m$, where the translation vector along the edge is the sum of $m$ zigzag vectors and one armchair vector. We call this a generalized zigzag edge, see Fig.~\ref{fig:super}.  The supercell defined by the lattice vectors, Eq.~\ref{eq:z6a}, contains $m+2$ unit cells. For instance, values $m=1,\,2$ give angles $\theta=79.1^{\mathrm{o}},\,70.9^{\mathrm{o}}$, respectively. The construction of the Hamiltonian matrix blocks defining the edge is straightforward. As an example, the Hamiltonian matrix blocks for $m=2$, $n=1$ are 
\begin{equation}
\label{eq:z7}
\mathbf{H}=\left(\begin{array}{cccc}
\mathbf{H}_{0,0} & \mathbf{H}_{1,1} & 0 & \mathbf{H}_{-1,0}e^{-i 2\pi k} \\
\mathbf{H}_{-1,-1} & \mathbf{H}_{0,0} & \mathbf{H}_{1,1} & 0 \\
0 & \mathbf{H}_{-1,-1} & \mathbf{H}_{0,0} & \mathbf{H}_{1,1} \\
\mathbf{H}_{1,0}e^{i 2\pi k} & 0 & \mathbf{H}_{-1,-1} & \mathbf{H}_{0,0}
\end{array}\right),
\end{equation}
\begin{equation}
\label{eq:z8}
\mathbf{B}=\left(\begin{array}{cccc}
\mathbf{H}_{0-1} & \mathbf{H}_{1,0} & 0 & \mathbf{H}_{-1,-1}e^{-i 2\pi k} \\
0 & \mathbf{H}_{0,-1} & \mathbf{H}_{1,0} & 0 \\
0 & 0 & \mathbf{H}_{0,-1} & \mathbf{H}_{1,0} \\
0 & 0 & 0 & \mathbf{H}_{0,-1}
\end{array}\right).
\end{equation}

To generate edges with an orientation closer to the armchair edge ($\theta=90^{\mathrm{o}}$) one can use the series with $m=1$ and vary $n$. The translation vector parallel to the edge is then the sum of one zigzag vector and $n$ armchair vectors, which we call a generalized armchair edge, see Fig.~\ref{fig:super}. The supercell then contains $2n+1$ unit cells. As an example, $m=1$, $n=3$ gives $\theta=85.3^{\mathrm{o}}$. The edge termination can be controlled by adding and removing atoms at the edge, and apply the technique outlined in Sec.~\ref{sec:modedge}.  

\subsection{Tight-binding models}
\label{sec:tbmodels}
We consider tight-binding models with spin degeneracy, i.e., we neglect spin-orbit coupling. The main contributions to the valence and conduction bands of \MX\ around the band gap come from the valence $d$-shell of the M atom and the $p$-shell of the X atoms. A minimal basis set then comprises eleven orbitals: five metal $d$-orbitals, and six $p$-orbitals of the two chalcogen atoms.\cite{Cappel:prb88} Monolayer \MX\ has $D_{3h}$ point-group symmetry. The trigonal prismatic coordination of the metal atom splits the \textit{d}-states into three groups: $A_{1}^{'}$\{$d_{3z^2-r^2}$\}, $E^{'}$\{$d_{xy},d_{x^2-y^2}$\}, and $E^{''}$\{$d_{xz},d_{yz}$\}. The six $p$-orbitals of the two chalcogen atoms split into the groups $A_{1}^{'}$\{$p_z^-=p_z(\mathrm{X}_1)-p_z(\mathrm{X}_2)$\}, $E^{'}$\{$p_x^+=p_x(\mathrm{X}_1)+p_x(\mathrm{X}_2),p_y^+=p_y(\mathrm{X}_1)+p_y(\mathrm{X}_2)$\}, $A_{2}^{''}$\{$p_z^+=p_z(\mathrm{X}_1)+p_z(\mathrm{X}_2)$\} and $E^{''}$\{$p_x^-=p_x(\mathrm{X}_1)-p_x(\mathrm{X}_2),p_y^-=p_y(\mathrm{X}_1)-p_y(\mathrm{X}_2)$\}. Mirror symmetry in the plane of the metal atoms, $\sigma_{h}$, allows for hybridization  between orbitals that are even with respect to $\sigma_{h}$, i.e., $A_{1}^{'}$ and $E^{'}$ orbitals, or between orbitals that are odd, i.e., $A_{2}^{''}$ and $E^{''}$ orbitals. 

The set of orbitals with even symmetry thus comprises the six orbitals $d_{3z^2-r^2}$, $d_{xy}$, $d_{x^2-y^2}$, $p_z^-$, $p_x^+$, and $p_y^+$, and the set with odd symmetry the five orbitals $d_{xz}$, $d_{yz}$, $p_z^+$, $p_x^-$, and $p_y^-$. As the even/odd symmetry is conserved for the edges and grain boundaries considered in this paper, all corresponding Hamiltonian matrices are blocked, and the even/odd solutions can be obtained separately. The matrix blocks $\mathbf{H}_{p,q}$ required for constructing the Hamiltonian matrices of Sec.~\ref{sec:edges}, are given in appendix~\ref{sec:11band}. The values of the tight-binding parameters have been obtained by fitting the bulk band structure to bands obtained from density functional theory (DFT) calculations with the generalized gradient approximation (GGA/PBE) functional.\cite{Perdew:prl96} For the even states we use the parameters given by Rostami \textit{et al.},\cite{Rostami:prb15} and for the odd states we use the parameters given in appendix~\ref{sec:11band}. 

The eleven band model can be simplified further. From early theoretical studies and recent first-principles calculations one observes that the \MX\ bands at the top of the valence band and at the bottom of the conduction band, are dominated by the metal \textit{d}-orbitals, in particular those with even symmetry, i.e., $d_{3z^2-r^2}$, $d_{xy}$, and $d_{x^2-y^2}$.\cite{Mattheis:prb73,Coehoorn:prb87,Cappel:prb88} Contributions to the bands around the gap from the $d_{xz}$ and $d_{yz}$ orbitals and from the chalcogen $p$-orbitals are much smaller. Matheiss has constructed an effective tight-binding model for \MX, where only the metal sites are taken into account explicitly.\cite{Mattheis:prb73} These sites form a two-dimensional triangular lattice, and the presence of the X atoms lowers the symmetry of this lattice from $D_{6h}$ to $D_{3h}$. The metal orbitals with even symmetry, $d_{3z^2-r^2}$, $d_{xy}$, and $d_{x^2-y^2}$, are used to construct an effective three-band model. The matrix blocks $\mathbf{H}_{p,q}$ of this model are given in appendix~\ref{sec:3band}. We use the parameters given by Liu \textit{et al.}\cite{Liu:prb13}, which have been obtained by fitting the bulk bands to GGA/PBE results. 

\begin{figure*}[t]
\includegraphics[width=2.0\columnwidth]{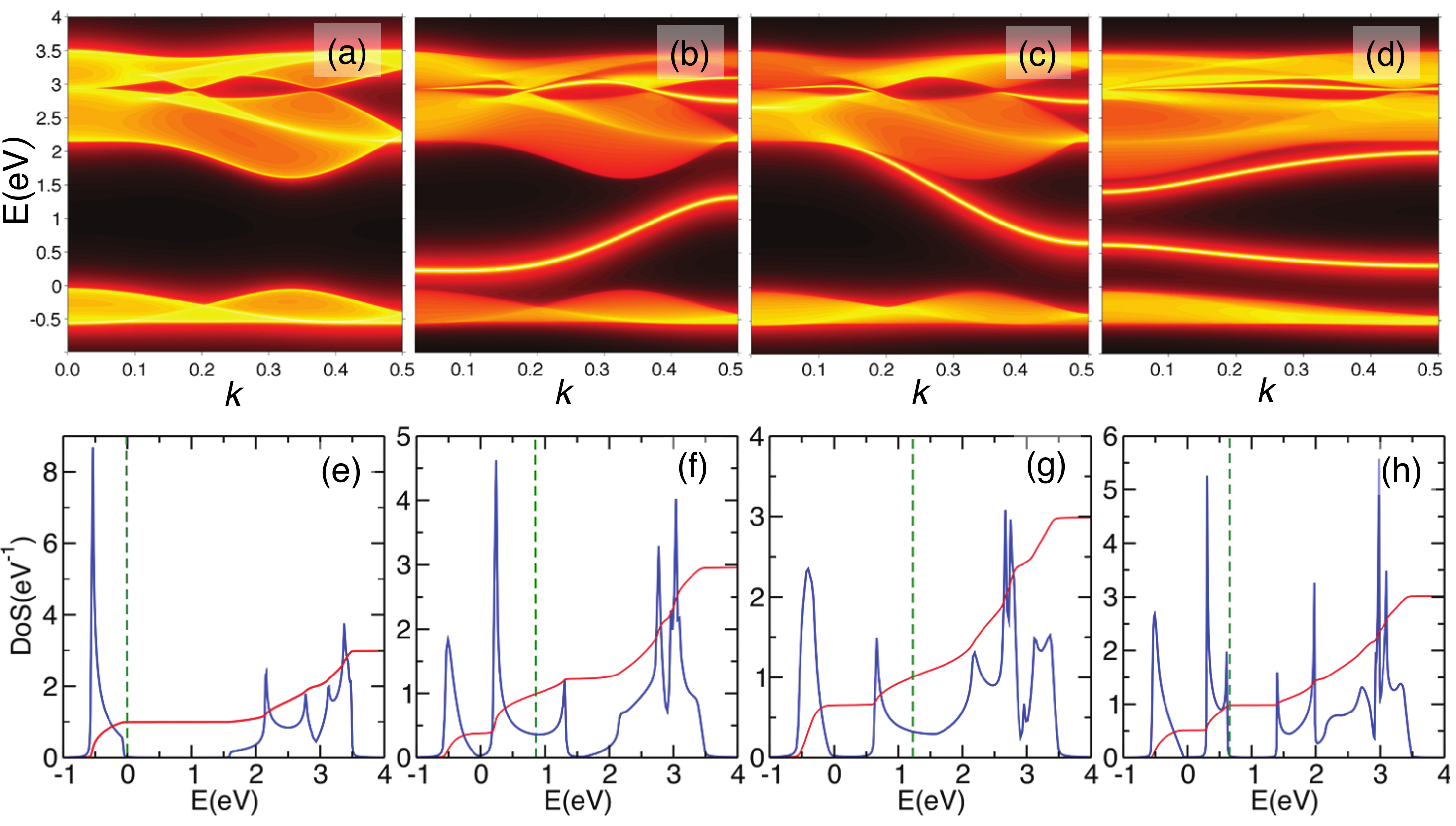}
\caption{(Color online) (a) $k$-resolved density of states (DoS) per \MS\ unit of bulk \MS\ in the three-band tight-binding model, with the $k$-vector parallel to the zigzag edge, and the zero of energy at the top of the valence band. The DoS is plotted on a logarithmic scale (right-hand side $n$ denotes amplitude $10^{-n}$) using a broadening parameter $\eta = 0.05$ eV (Eq.~\ref{eq:g1}). (b), (c), (d) $k$-resolved DoSs of the Mo-edge, the S-edge, and the armchair edge, respectively. (e),(f),(g),(h) $k$-integrated DoS per \MS\ unit of bulk \MS, the Mo-edge, the S-edge, and the armchair edge, all plotted on a linear scale ($\eta = 0.05$ eV). The red solid lines give the counting function and the green dashed lines indicate the charge neutrality level (CNL), Eqs.~\ref{eq:s8}-\ref{eq:s10}.
}
\label{fig:zigzag}
\end{figure*}

\section{Results}
\subsection{Three-band model}
\label{sec:result1}
\subsubsection{Zigzag and armchair edges}

The eigenvalues and eigenvectors of Eq.~\ref{eq:e3a}, are used to construct the Bloch matrices, Eq.~\ref{eq:e6}. The bulk density of states $n(k,E)$ is shown in Fig.~\ref{fig:zigzag}(a), as calculated from the Green's function matrix of a bulk strip, Eqs.~\ref{eq:g2} and \ref{eq:b5}, in the zigzag edge orientation. In the three-band model, the lowest band is the valence band, and the overlapping two upper bands form the conduction band. Note that the zero of energy is chosen at the top of the valence band. The same Bloch matrices give access to the Green's function matrices of the zigzag edge strips, Eqs.~\ref{eq:b3} and \ref{eq:b4}. Here $\mathbf{g}_R(k,E)$ and $\mathbf{g}_L(k,E)$ are the energy and $k$-resolved Green's function of the M-edge and the X-edge, respectively, see Fig.~\ref{fig:lat}. Again using \MS\ as an example, the associated densities of state $n_R(k,E)$ and $n_L(k,E)$ of the Mo-edge, respectively the S-edge, are shown in Fig.~\ref{fig:zigzag}(b) and (c).     

The Mo-edge has a prominent edge state inside the bulk band gap dispersing upwards from $k=0$ to $k=\frac{1}{2}$. The S-edge has an edge state that starts in the bulk conduction band at $k=0$, and disperses downward to $k=\frac{1}{2}$. In surface physics such states are termed Shockley states.\cite{Pershoguba:prb12} These edge states are also found in the more detailed eleven-band model with a similar dispersion, see Sec.~\ref{sec:result2}. The latter model shows a richer edge state structure, originating from bands that are omitted in the simple three-band model. Nevertheless, the three-band model finds prominent Mo-edge and the S-edge states with the correct dispersion. Edge states are also found at energies close to $\sim 3$ eV, which is in the hybridization gap between the two conduction bands.   

The structure of a bulk strip in armchair orientation has a mirror plane perpendicular to the \MX-plane and along the armchair orientation, see see Fig.~\ref{fig:lat}. It follows that the densities of states of right and left edges, $n_R(k,E)$ and $n_L(k,E)$, are identical for the armchair orientation. The $k$-resolved density of states of the \MS\ armchair edge, Eq.~\ref{eq:z3}-\ref{eq:z4}, is shown in Fig.~\ref{fig:zigzag}(d). There are two clear edge states with energies in the band gap. One edge state is just below the conduction band and roughly follows the dispersion of the conduction band edge, whereas the other one is positioned at $\sim 0.5$ eV above the valence band, following the dispersion of valence band edge. 

The $k$-integrated densities of states, Eq.~\ref{eq:s10}, of a bulk \MS\ strip in zigzag orientation, the Mo-edge, the S-edge, and the armchair edge, are given in Figs.~\ref{fig:zigzag}(e)-(h), respectively. These densities of states show the van Hove singularities at the band edges that are typical of 1D structures. With the zero of energy at the top of the valence band, the bulk valence band lies in the energy range $-0.5$-0.0 eV, and the two conduction bands in the range 1.6-3.5 eV, see Fig.~\ref{fig:zigzag}(e). The Mo-edge shows edge bands in the energy range 0.1-1.4 eV, and around 3 eV. The S-edge has an edge band starting at 0.5 eV, which merges with the conduction band at higher energies, and additional edge states with energies 2.7-3.0 eV. The armchair edge has two edge band in the gap at in the energy ranges 0.3-0.6 eV and 1.4-2.0 eV, respectively, and an edge state around 3.1 eV.

Also shown in Figs.~\ref{fig:zigzag}(e)-(h) are the electron counting function  (red curves), Eq.~\ref{eq:s9}, and the charge neutrality level (CNL; green lines), Eq.~\ref{eq:s8}. Obviously for the bulk the CNL is at the top of the valence band, see Fig.~\ref{fig:zigzag}(e). The position of the CNL at the Mo-edge corresponds to a $2/3$ filled edge band in the gap, see Fig.~\ref{fig:zigzag}(f). One might interpret this as effectively two-third of the bonds being broken for a Mo atom at the Mo-edge. Likewise, the position of the CNL at the S-edge corresponds to a $1/3$ filled edge band in the gap, see Fig.~\ref{fig:zigzag}(g). This correlates with one-third of the bonds being broken for a Mo atom at the S-edge. The supercell of the armchair edge has two Mo atoms along the edge, where the local environments of one is similar to that of a Mo atom at an Mo-edge, and the local environment of the other is similar to that of a Mo atom at the S-edge, see Fig.~\ref{fig:lat}. Summing the $2/3$ and $1/3$ edge state occupations at the Mo- and S-edges, one predicts that  the armchair edge has one completely filled edge band. The calculated CNL shows that this is indeed the case; the armchair edge has one fully occupied edge band, and one empty edge band, see see Fig.~\ref{fig:zigzag}(h).   

The CNLs of all three basic edges are quite close, cf. Figs.~\ref{fig:zigzag}(f)-(h). This is an artefact of the three-band model, and of disregarding the odd bands in particular. The latter play an important role in setting the CNLs in \MX\ edges, as we will discuss in Sec~\ref{sec:result2}. In polar lattices such as \MX\ the CNL does not fix the intrinsic Fermi level, unlike in a non-polar lattice such as graphene.\cite{Guller:prb13,Gilbertini:nl15} In a finite-sized \MX\ sample electrons can be redistributed among all the edges in the sample, driven by the internal electric field set up by the intrinsic polarization of the material. 

\begin{figure}[t]
\includegraphics[width=8.5cm]{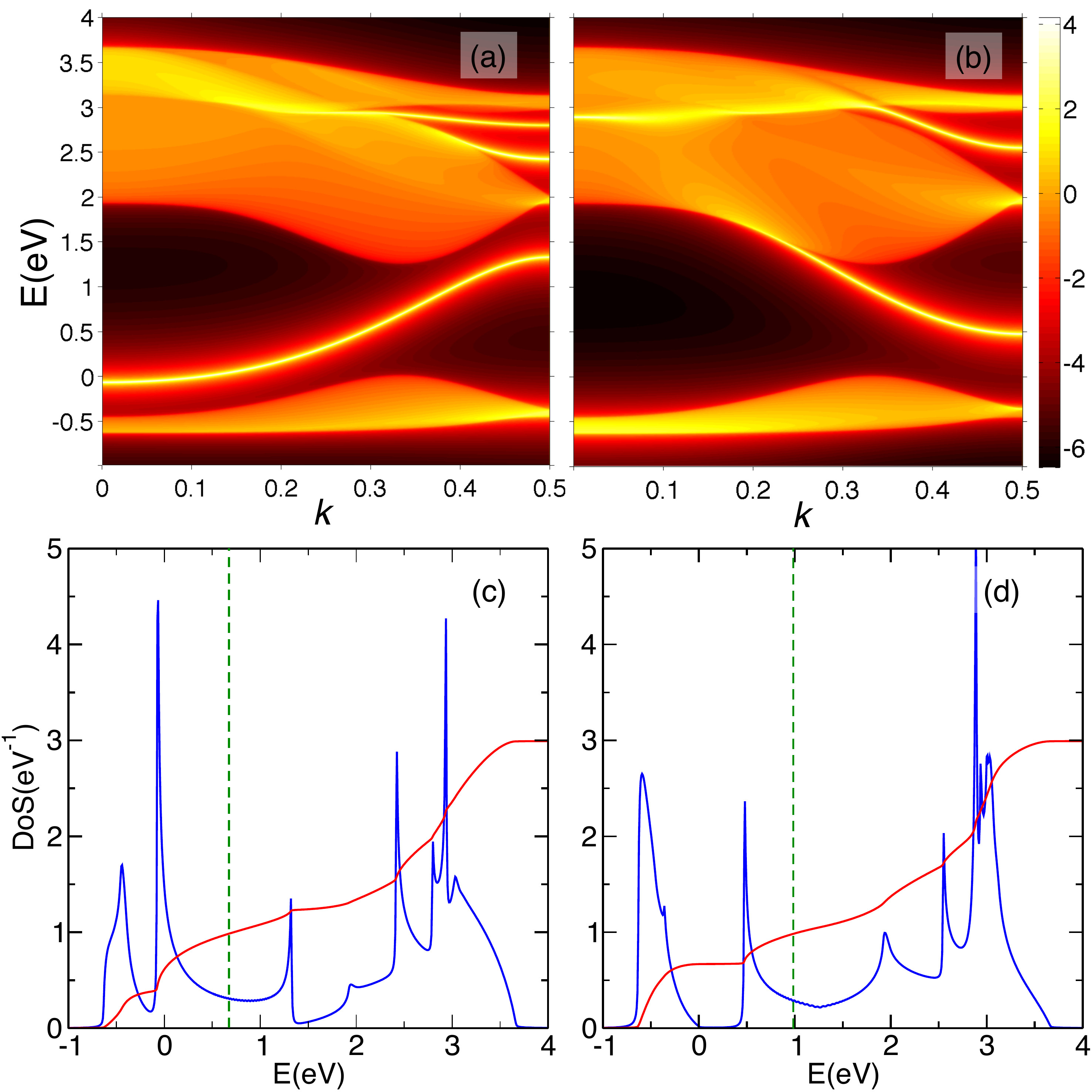}
\caption{(Color online) (a),(b) $k$-resolved DoSs of the W-edge and the Te-edge within the three band tight-binding model of $H$-\WT. (c),(d) $k$-integrated DoSs of the W-edge and the Te-edge. The same conventions and parameters are used as in Fig.~\ref{fig:zigzag}.}
 \label{fig:WTe2}
\end{figure}

So far we have focused on edges of \MS, but all \MX\ compounds with a similar structure ($D_{3h}$ point group, trigonal prismatic coordination of M atoms by X atoms) have similar edge structures. Figure~\ref{fig:WTe2} shows the densities of states of the W and Te zigzag edges of $H$-\WT\ as an example. These densities of states are very similar to those of the corresponding Mo- and S-edges of \MS, see Figs.~\ref{fig:zigzag}(b), (c), (f) and (g). The band gap of bulk \WT\ is somewhat smaller than that of \MS, and the band widths are somewhat larger. \WT\ has a band gap of 1.1 eV (GGA/PBE), a valence band in the range $-0.7$-0.0 eV, and conduction bands in the range 1.1-3.7 eV. The W-edge has a state in the gap in the energy range 0.0-1.4 eV, and the Te-edge has a state in the gap in the range 0.5-2.0 eV. As is the case for the bulk bands, the band widths of these edge states are somewhat larger than the corresponding states in \MS. 

\subsubsection{General edges and grain boundaries}

\begin{figure}[t]
\includegraphics[width=8.5cm]{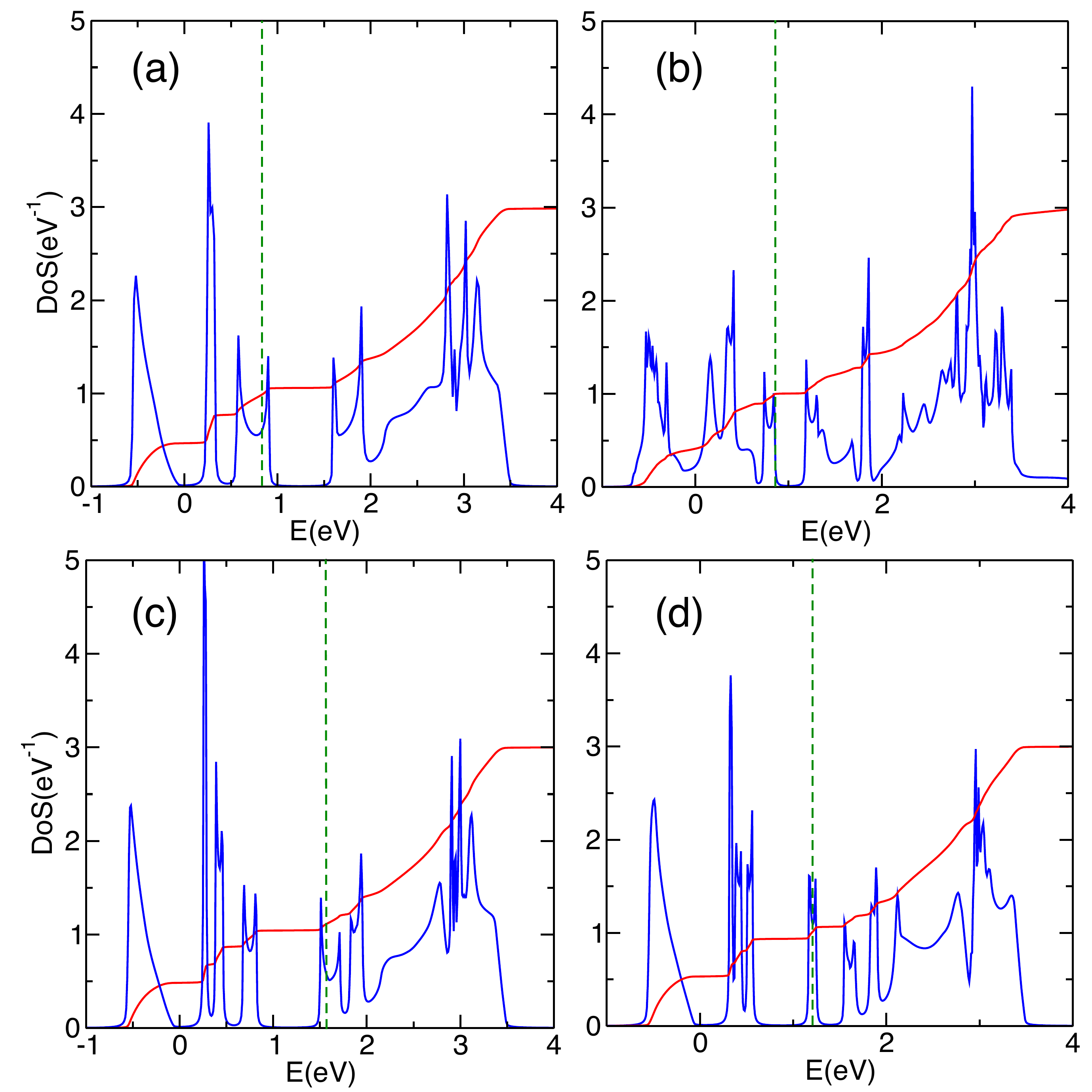}
\caption{(Color online) $k$-integrated DoSs of the generalized zigzag edges defined by Eq.~\ref{eq:z6a} with (a) $m=1,n=1$ and (b) $m=3,n=1$, and of the generalized armchair edges with (c) $m=1,n=2$ and (d) $m=1,n=3$.}
\label{fig:genedge}
\end{figure}

The electronic structure of the more general edges defined in Sec.~\ref{sec:genedge} can be calculated along the same lines. Figs.~\ref{fig:genedge}(a) and (b) give the densities of states of generalized zigzag edges of \MS\ with translation vectors along the edge defined by $m=1$ and 3, respectively, and $n=1$, see Eq.~\ref{eq:z6a} and Fig.~\ref{fig:super}. The generalized zigzag edge has a rich structure of edge states within the bulk band gap. The peak at 0.4 eV in Figs.~\ref{fig:genedge}(a) and (b) results from the armchair part of this edge, compare to Fig.~\ref{fig:zigzag}(h). The band in the range 1-1.5 eV in \ref{fig:genedge}(a), and the structure of bands in the range 0.8-1.9 eV in \ref{fig:genedge}(b) originates mainly from the zigzag parts. 

The counting system for the three-band model, as outlined in the previous section, gives a filling 2/3 per Mo atom at a Mo-edge and a filling 1 per two Mo atoms at an armchair edge. For a generalized zigzag edge one would then predict the CNL to correspond to $2m/3+n$ filled edge states. This would mean that the CNL is inside an edge band, unless $m$ is a multiple of three. This is confirmed by a calculation of the CNL according to Eq.~\ref{eq:s8}. In Fig.~\ref{fig:zigzag}(a), where $m=1$, the position of the CNL corresponds to a 2/3-filled edge band, whereas in Fig.~\ref{fig:zigzag}(b), where $m=3$, the CNL is in a gap between two edge states. The gaps between the edge states of the general edges are quite small however. In a sample that involves long range charge transfer between the edges, as discussed above, these semiconducting edges easily become doped.

Figures~\ref{fig:genedge}(c)and (d) show the densities of states of generalized armchair edges of \MS, with translation vectors along the edge with  $m=1$ and $n=2$ and 3, respectively, see Eq.~\ref{eq:z6a} and Fig.~\ref{fig:super}. Again this general edge has a rich structure of states within the bulk band gap. In particular the peaks around 0.4 eV and 1.7 eV have a strong armchair character. The counting model gives $1/3+n$ filled edge states, so it predicts that the CNL always lies inside an edge band. Calculations of the CNL according to Eq.~\ref{eq:s8} confirm this, as shown in Fig.~\ref{fig:genedge}(c) and (d). In conclusion, within the three-band tight-binding model, pristine charge neutral edges are metallic for most edge orientations. 

\begin{figure}[t]
\includegraphics[width=8.5cm]{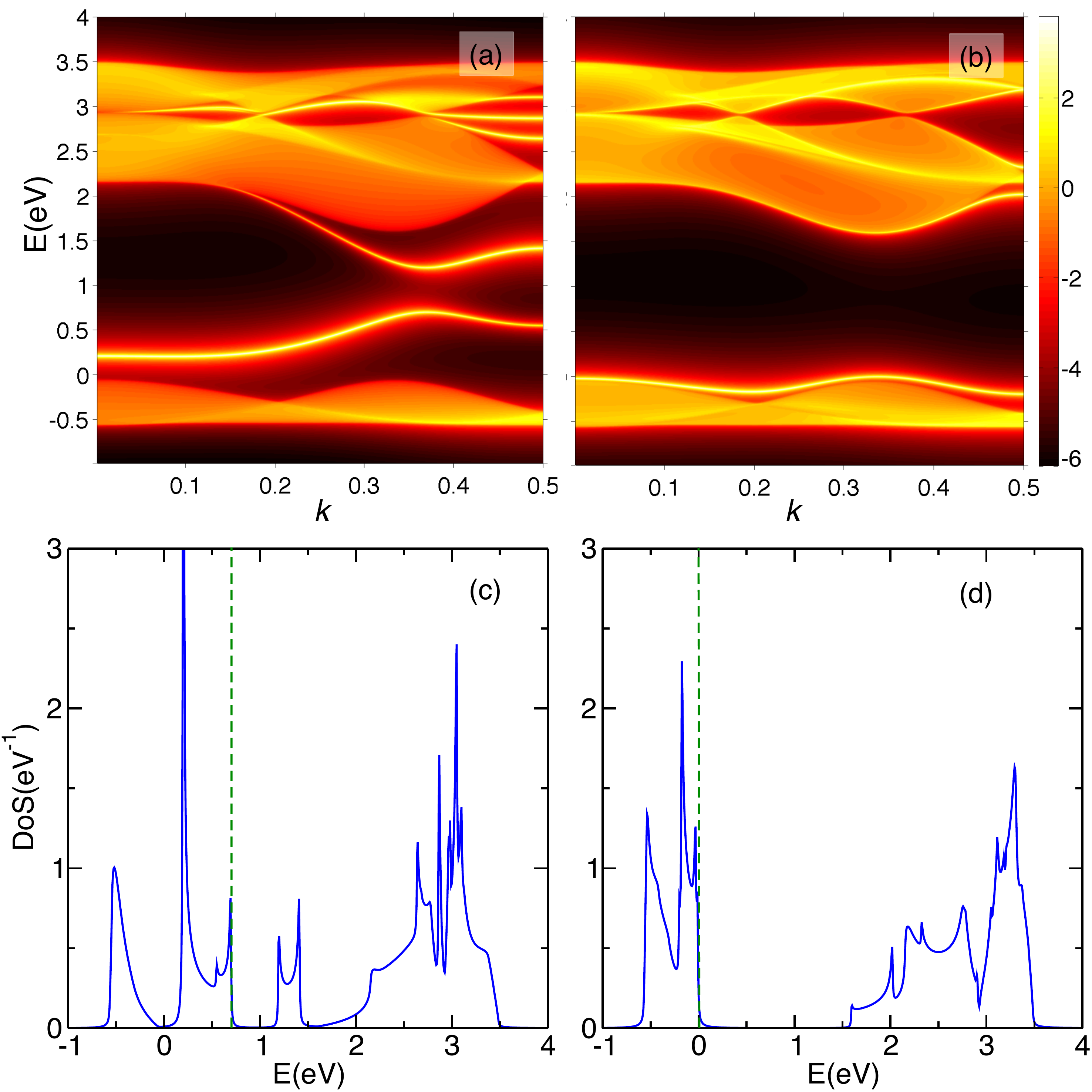}
\caption{(Color online) (a) $k$-resolved DoS of a model grain boundary between a Mo-edge and a S-edge, with a coupling between the edges $\alpha=0.2$, Eq.~\ref{eq:z1a}. (b) $k$-resolved DoS with $\alpha=0.8$. (c),(d) the corresponding $k$-integrated DoSs.}
\label{fig:boundary}
\end{figure}

A similar analysis can be applied to the states found at grain boundaries. We illustrate the use of Eq.~\ref{eq:s3} for calculating the states at grain boundaries using a simple grain boundary in \MS. It consists of a left semi-infinite part, terminated by a S-edge, connected to a right semi-infinite part, terminated by a Mo-edge. As hopping matrix defining the coupling between left and right parts,  Fig.~\ref{fig:grainboundary}, we choose a scaled version of the bulk hopping matrix, Eq.~\ref{eq:z1a}.
\begin{equation}
\mathbf{B'}=\alpha \mathbf{B};\; 0\leq \alpha \leq 1.
\end{equation}
Obviously $\alpha = 0$ gives two uncoupled S- and Mo-edges with corresponding edge states, Figs.~\ref{fig:zigzag}(b), (c), (f) and (g), whereas $\alpha = 1$ gives the bulk electronic structure, Figs.~\ref{fig:zigzag}(a) and (e), without any edge states. The values $0 < \alpha < 1$ then represent a simple model for a weak link with zigzag orientation.

Figure~\ref{fig:boundary}(a) gives the $k$-resolved density of states for $\alpha=0.2$, which represents a relatively weak coupling between the left and right parts. In the band gap one observes the two edge bands that are typical of the Mo-edge and the S-edge, see Figs.~\ref{fig:zigzag}(b) and (c). Due to the coupling between the S- and Mo-edges at the grain boundary, the two bands interact, which results in an avoided crossing between the two edge states at $k\approx0.35$ and $E\approx 1$ eV. The avoided crossing creates an energy gap in the range 0.7-1.2 eV, which is clearly visible in the $k$-integrated density of states for $\alpha=0.2$, shown in Fig.~\ref{fig:boundary}(c). The counting model predicts 2/3, respectively 1/3 filling per Mo atom of an edge state at the Mo-edge and the S-edge, if the grain boundary is charge neutral. This implies that the lower edge band is fully occupied, whereas the upper edge band is empty. A calculation of the CNL according to Eq.~\ref{eq:s8} confirms this, see Fig.~\ref{fig:boundary}(c). 

\begin{figure}[t]
\includegraphics[width=1.0\columnwidth]{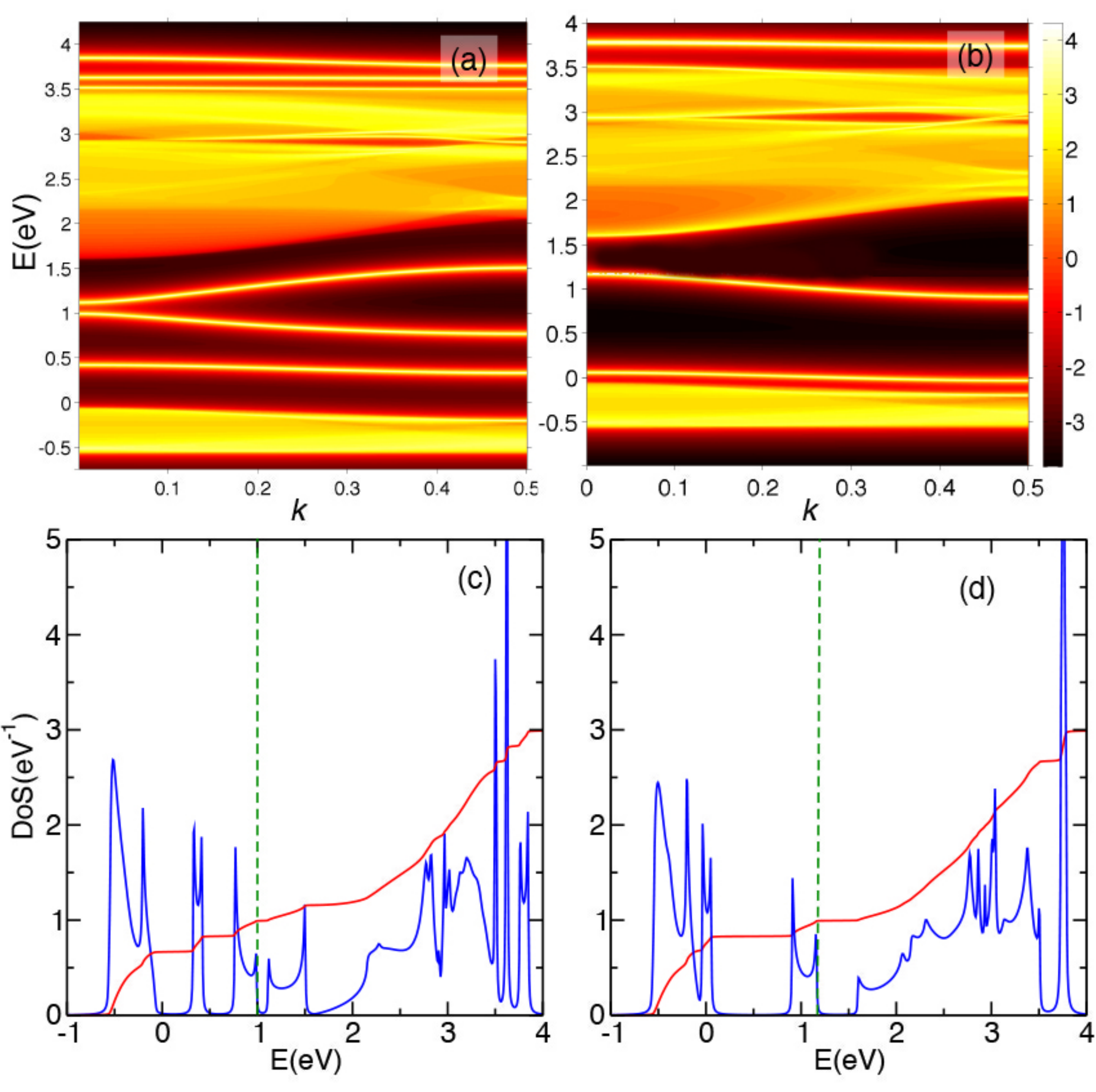}
\caption{(Color online) (a,c) $k$-resolved and $k$-integrated DoS of a Mo-edge with a $3\times$ reconstruction, where the on-site energy of every third Mo-atom is 1 eV lower than that of the other two; (b,d) the same information for the S-edge.}
\label{fig:triple}
\end{figure}

\begin{figure*}[t]
\includegraphics[width=1.9\columnwidth]{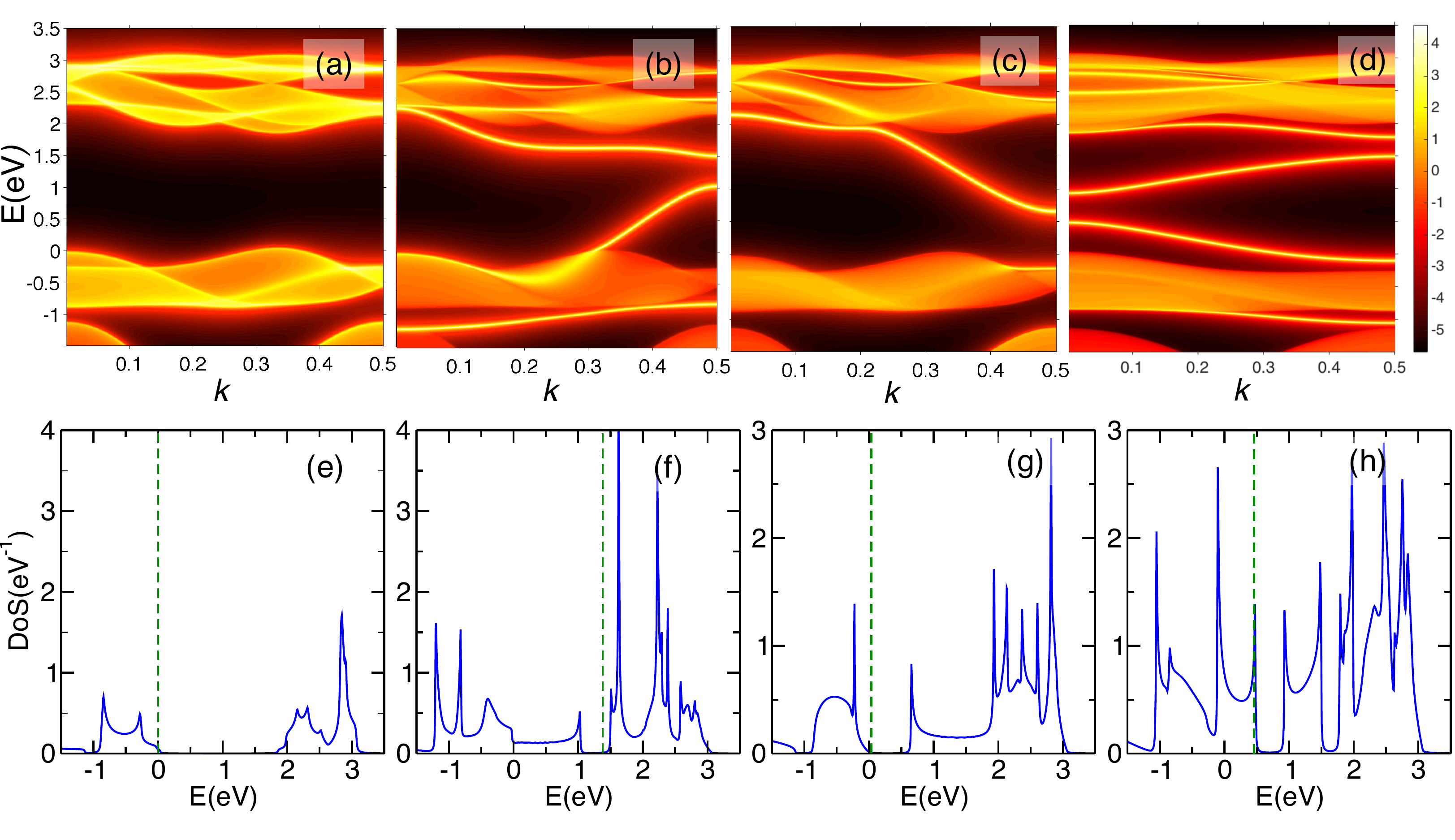}
\caption{(Color online) As Fig.~\ref{fig:zigzag} but for the states within the eleven band model with even symmetry with respect to a mirror plane through the plane of the Mo atoms. (a),(b),(c),(d) $k$-resolved DoS per \MS\ unit of bulk \MS, the Mo-edge, the S-edge, and the armchair edge, respectively. (e),(f),(g),(h) The corresponding $k$-integrated DoSs of bulk \MS, the Mo-edge, the S-edge, and the armchair edge. The same conventions and parameters are used as in Fig.~\ref{fig:zigzag}. 
}
 \label{fig:even11band}
\end{figure*}

Figure~\ref{fig:boundary}(b) gives the $k$-resolved density of states for a strong coupling between the left and right parts, $\alpha = 0.8$. The interaction between the two edge bands is now much stronger than for the case discussed in the previous paragraph, which results in a much larger gap in the range  0.2-1.6 eV, see Fig.~\ref{fig:boundary}(d). The two edge bands are pushed toward the valence band and the conduction band, respectively, and they more closely follow the dispersions of the valence and conduction band edges. The occupancy of a charge neutral grain boundary is the same as before, i.e., a fully occupied lower edge band and an empty upper edge band.

As a final point in this section we illustrate the technique introduced in Sec.~\ref{sec:modedge} for handling edges that involve an electronic or a structural reconstruction. For the zigzag edges we have found a 2/3 respectively a 1/3 occupied edge band for the Mo-edge and the S-edge. It suggests that a reconstruction that triples the translational period along the edge, may lead to fully occupied edge states for both edges. As a proof of principle, we test a very simple reconstruction, where one in three Mo atoms at the Mo-edge or the S-edge has a different on-site energy. The Green's function matrix is calculated from Eq.~\ref{eq:s5}. The densities of states of the modified edge layers are shown in Fig.~\ref{fig:triple}(a) and (c) for the Mo-edge and (b) and (d) for the S-edge. Obviously increasing the periodicity from $1\times$ to $3\times$ folds the edge band, so one observes three edge states instead of one. Decreasing the on-site energy of every third Mo atom at the edge by 1 eV introduces energy gaps between the edge states. 

By calculating the CNL we observe that at the Mo-edge the two lowest edge bands, starting at 0.4 eV and 0.7 eV in Fig.~\ref{fig:triple}(b) and (d), are filled, whereas the third band, starting at 1.1 eV, is empty. The lowest edge band of the S-edge at 0.0 eV, Fig.~\ref{fig:triple}(a) and (c), is filled, whereas the two upper bands, starting at 0.9 eV and 1.6 eV, respectively, are empty.

\subsection{Eleven-band tight-binding model}
\label{sec:result2}

\begin{figure*}[tb]
\includegraphics[width=1.4\columnwidth]{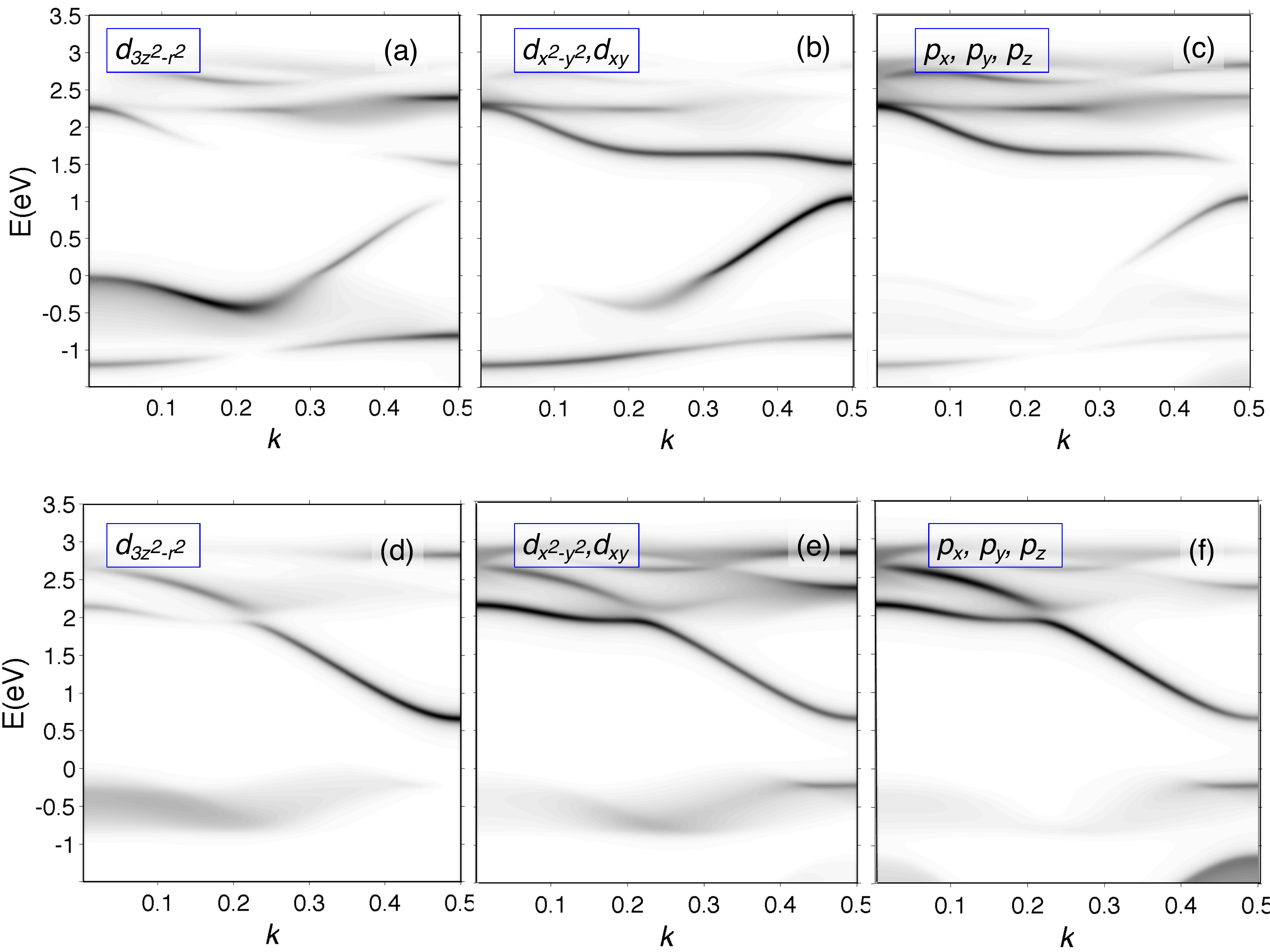}
\caption{(Color online) Projected densities of states (PDoS) calculated according to Eq.~\ref{eq:g3a}. (a),(b),(c) For the Mo-edge, projected on the Mo $d_{3z^2-r^2}$ orbital, the $d_{xy}$ and $d_{x^2-y^2}$ orbitals, and the $p$ orbitals of the S atoms, respectively. (d),(e),(f) The same information for the S-edge. The PDoSs are calculated using a broadening parameter $\eta=0.1$ eV, Eq.~\ref{eq:g1}, and are plotted with a linear grayscale.}
 \label{fig:proj11band}
\end{figure*}

As discussed in Sec.~\ref{sec:tbmodels}, the eleven-band tight-binding model of \MX\ uses a basis set composed of the five valence $d$-orbitals of the M atom, and the six valence $p$ orbitals of the two X atoms. Because mirror symmetry $\sigma_h$ in the \MX\ plane holds both for monolayers as well as for edges, states that are even or odd with respect this mirror symmetry are treated separately. The even symmetry set comprises the six orbitals $d_{3z^2-r^2}$, $d_{xy}$, $d_{x^2-y^2}$, $p_z^-$, $p_x^+$, and $p_y^+$, and the odd symmetry set the five orbitals $d_{xz}$, $d_{yz}$, $p_z^+$, $p_x^-$, and $p_y^-$, leading to six- and five-dimensional Hamiltonian matrices, respectively. 

Again we use \MS\ as an example of a \MX\ monolayer. The procedure for obtaining the Green's functions and the densities of states are the same as those described in the previous section. In the following we will only discuss the results obtained for the basic edges, the zigzag and the armchair edges, and compare those to the results obtained with the three-band model.
 
\subsubsection{Even bands} 
 
The $k$-resolved density of states of the even states of a bulk strip in the zigzag edge orientation is shown in Fig.~\ref{fig:even11band}(a) in an energy region around the band gap, and the corresponding $k$-integrated density of states is given in Fig.~\ref{fig:even11band}(e). As before, the zero of energy is positioned at the top of the valence band. Qualitatively, these densities of states are similar to those of the three-band model, compare to Figs.~\ref{fig:zigzag}(a) and (e). Quantitatively, the eleven-band model gives a highest valence band that is wider by $\sim 0.3$ eV, whereas the two lowest conduction bands are narrower by $\sim 0.5$ eV in total. 

Figures~\ref{fig:even11band}(b) and (f) show the $k$-resolved and $k$-integrated densities of states of the Mo-edge. There are two prominent edge bands with energies in the \MS\ band gap. One edge band emerges from the bulk valence band at $k\approx 0.3$, and disperses upward with increasing $k$ to 1 eV. A second band disperses downward from the conduction band to 1.6 eV. The first edge band is also found in the three-band model, compare Figs.~\ref{fig:zigzag}(b) and (f). In the three-band model it is found at a slightly higher energy, such that it is completely isolated from the bulk valence band. The second edge band is absent in the three-band model. We will discuss the character of these bands in  more detail below. Like in the three-band model, there are also edge states at other energies, for instance just below the highest valence band at $\sim-1.1$ eV, and in the hybridization gaps of the conduction bands.  

The $k$-resolved and $k$-integrated densities of states of the S-edge are given in Figs.~\ref{fig:even11band}(c) and (g). At energies in the \MS\  band gap there is one prominent edge state, which starts at $k=0$ in the conduction band, and disperses downward with increasing $k$ to 0.7 eV. The three-band model shows the same edge state with more or less the same dispersion, compare to Figs.~\ref{fig:zigzag}(c) and (g). In the eleven-band model this state is somewhat more prominently isolated from the conduction band. Unlike for the Mo-edge, the eleven-band model does not give additional even edge states for the S-edge in the band gap, as compared to the three-band model. 

Finally, Figs.~\ref{fig:even11band}(d) and (h) show the $k$-resolved and $k$-integrated densities of states of the armchair edge. There are three edge bands with energies in the \MS\ band gap. One band is situated at $-0.1$-0.5 eV and a second edge band lies at 0.9-1.5 eV. These two edge bands roughly correspond to the ones that are found in the three-band model of the armchair edge, see Figs.~\ref{fig:zigzag}(d) and (h). In the eleven-band model these two edge bands are found at a slightly lower energy in the gap. In addition  the present model finds a third edge band in the gap, just below the conduction band at 1.7-1.9 eV.

\begin{figure*}[tbh]
\includegraphics[width=2.0\columnwidth]{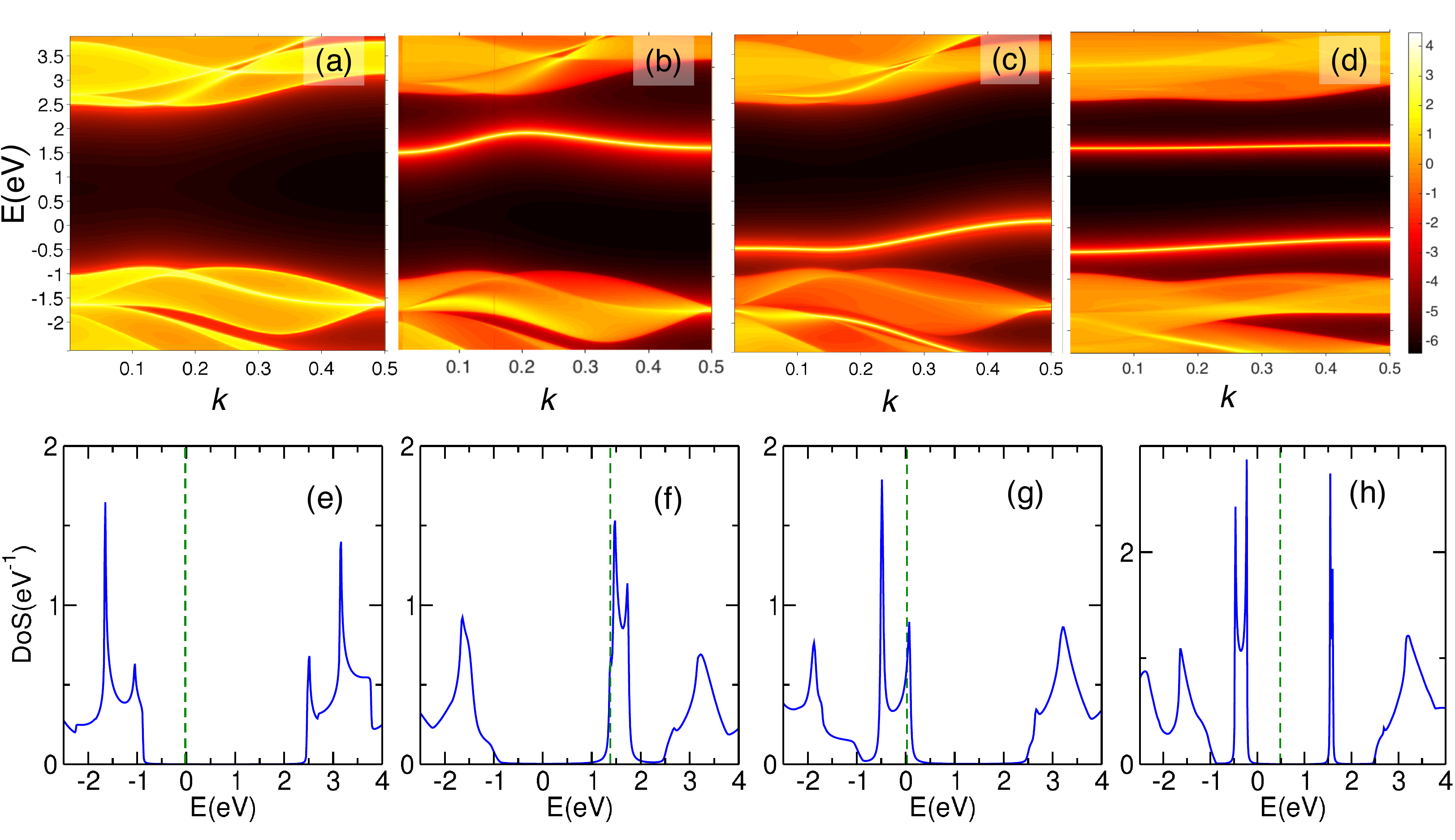}
\caption{(Color online) As Fig.~\ref{fig:even11band}, but for the states with odd symmetry with respect to a mirror plane through the plane of the Mo atoms.}
 \label{fig:odd11band}
\end{figure*}

In conclusion, although there are quantitative differences between the electronic structures found with the eleven-band model and the three-band model, qualitatively they give similar results concerning the prominent edge states of even symmetry found in the \MS\ band gap. For the Mo-edge and the armchair edge, the eleven-band model gives an additional edge state, as compared to the three-band model, with energies just below the conduction band.

Figures~\ref{fig:even11band}(e)-(h) also show the CNLs. One needs of course all the bands to calculate the CNLs, including the odd bands to be discussed in the next section. A comparison to Fig.~\ref{fig:zigzag} reveals that the odd bands are in fact instrumental in fixing the CNLs. For instance, the CNLs of the Mo-edge and the S-edge differ by 1.4 eV, Figs.~\ref{fig:even11band}(f) and (g), whereas the corresponding CNLs in Figs.~\ref{fig:zigzag}(f) and (g) differ by 0.3 eV only. The eleven-band model gives one completely occupied even edge band for the Mo-edge and one empty one, see Fig.~\ref{fig:even11band}(f). For the S-edge it gives one completely empty even edge band, see Fig.~\ref{fig:even11band}(g). This is unlike the the three-band model, where we found partially filled even edge bands, both for the Mo-edge and the S-edge. As we will see in the next section, edge states of odd symmetry, which are absent in the three-band model, pin the CNLs of the Mo-edge and the S-edge, and make these edges metallic. The CNL at the armchair edge is in the gap between the two edge bands, see Fig.~\ref{fig:even11band}(h), as it also is in the three-band model, see Fig.~\ref{fig:zigzag}(h)

The orbital character of the edge states can be analyzed using the projected density of states, according to Eq.~\ref{eq:g3a}. The projections on the Mo $d_{3z^2-r^2}$ orbital, the $d_{xy}$ and $d_{x^2-y^2}$ orbitals, and the $p$ orbitals of the S atoms are given in Figs.~\ref{fig:proj11band}(a), (b), and (c), respectively, for the Mo-edge. The results indicate that both edge states in the \MS\ band gap have a mixed Mo and S character. The dominant Mo contributions clearly come from the $d_{xy}$ and $d_{x^2-y^2}$ orbitals. The upper edge state has S $p$ character mixed in, in particular for $k < 0.4$, whereas the lower edge state has some S $p$ character mixed in for $k > 0.4$. There is little $d_{3z^2-r^2}$ character mixed in these  edge bands, except at the band edges. 

The projected densities of states of the S-edge, projected on the same orbitals, are given in Figs.~\ref{fig:proj11band}(d), (e), and (f). Also here the edge state in the \MS\ band gap has a mixed Mo and S character. As for the Mo contribution, for $k=0$ the dominant character is Mo $d_{xy}$ and $d_{x^2-y^2}$. That changes somewhat for larger $k$; at $k=0.5$ the dominant character is Mo $d_{3z^2-r^2}$. The contribution of the S-orbitals is fairly constant throughout the whole edge band. 

\subsubsection{Odd bands}
The $k$-resolved and $k$-integrated bulk densities of states $n(k,E)$ corresponding to the odd states are given in Figs.~\ref{fig:odd11band}(a) and (e). The band gap between the odd states is 3.3 eV, which is significantly larger than the gap between the even states. The top of the highest valence band of the odd states is approximately 0.8 eV below the top of the highest valence band of the even states, whereas the bottom of the lowest conduction band of the odd states is approximately 0.6 eV above that of the even states.  

The $k$-resolved densities of state of the odd bands are shown in Figs.~\ref{fig:odd11band}(b) and (c) for the Mo-edge and of the S-edge, respectively, and the corresponding $k$-integrated densities of states are shown in Figs.~\ref{fig:odd11band}(f) and (g). Both edges have a prominent edge band with moderate dispersion inside the gap between the odd states. The edge band of the Mo-edge lies close to the conduction band between 1.3 and 1.8 eV, whereas the edge band of the S-edge is close to the top of the valence band between $-0.5$ and 0.2 eV. 

\begin{figure}[tb]
\includegraphics[width=1.0\columnwidth]{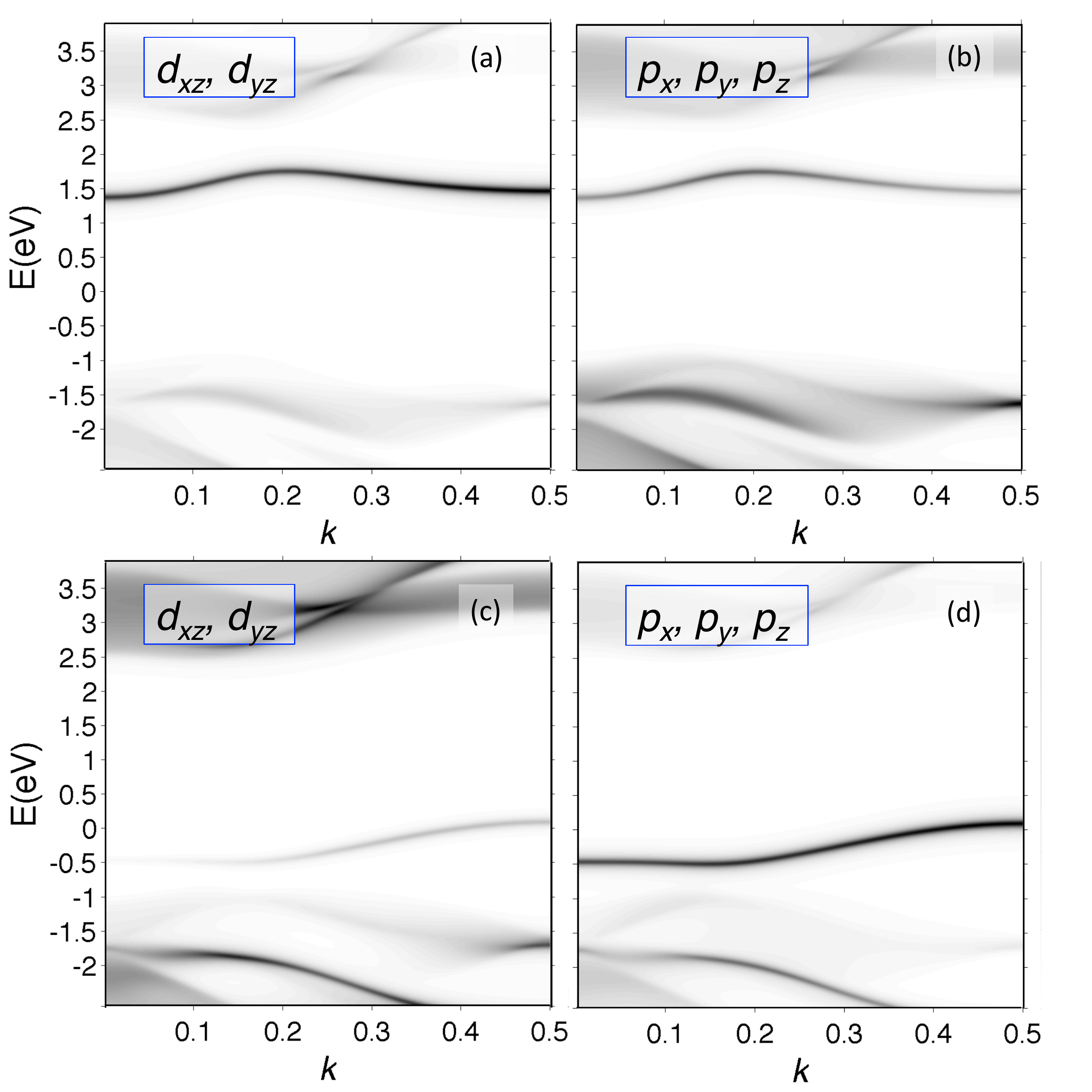}
\caption{(Color online) As Fig.~\ref{fig:proj11band}, but for the states with odd symmetry with respect to a mirror plane through the plane of the Mo atoms.} 
 \label{fig:projodd11band}
\end{figure}

If we also take the bulk band structure of the even bands into consideration, Figs.~\ref{fig:even11band}(a) and (d), then most of the edge band at the S-edge, Figs.~\ref{fig:odd11band}(c) and (g) overlaps with the valence band. It is still a true edge state though, because the interaction between the odd edge state and the even bulk bands is symmetry forbidden. The edge band at the Mo-edge, Figs.~\ref{fig:odd11band}(b) and (f) partially overlaps with the conduction band of the even states. Also this is a true edge state over the whole 1D Brillouin zone, as interaction with the bulk states is symmetry forbidden. 

Figures~\ref{fig:odd11band}(d) and (h) give the $k$-resolved and $k$-integrated densities of states corresponding to the armchair edge. These results show two edge bands in the gap region with a modest to small dispersion. The lowest of these edge bands has a dispersion of 0.4 eV, and taking also the even bulk bands into consideration, it lies in the \MS\ valence band. The upper edge band is near dispersionless; it is found in the gap at 1.6 eV. 

The calculated CNLs are also given in Figs.~\ref{fig:odd11band}(e)-(h). From these figures it becomes clear that the odd edge states are instrumental in controlling the CNLs of the edges. The CNL of the Mo-edge is found in the edge band that is close to the \MS\ conduction band, see Fig.~\ref{fig:odd11band}(f), and the CNL of the S-edge lies in the edge band close to the \MS\ valence band, see Fig.~\ref{fig:odd11band}(g). Within the eleven-band model both these edges are metallic in their charge neutral state. The CNL of the armchair edge lies between the two edge bands, Fig.~\ref{fig:odd11band}(h), which means that the neutral armchair edge is semiconducting.

The orbital character of the odd symmetry edge states is identified using the projected densities of state. Figures ~\ref{fig:projodd11band}(a) and (b) show the projections on the $d$- and $p$-orbitals of the density of states of the Mo-edge, and Figs.~\ref{fig:projodd11band}(c) and (d) show the projections for the S-edge. The edge state at the Mo-edge has somewhat of a mixed character, but it is dominated by Mo $d$-orbitals. The edge state at the S-edge has a clear S $p$ character. More than the even edge states discussed in the previous section, these two odd edge states have a ``dangling-bond'' character, i.e., they correspond to bulk bonds that are broken at the edge atoms.

\section{Summary and Conclusions}
Growth of 2D semiconductors, such as the transition metal dichalcogenides \MX, commonly results in sheets that contain quasi-1D metallic structures, which are located at the edges of 2D crystallites, or at the boundaries between 2D crystal grains. Such edges and grain boundaries form a practical realization of 1D metals. Metalic edges have also been identified as the active sites in \MX\ catalysts. A standard technique for modeling edge or grain boundary states starts from nanoribbon structures, which involves using large supercells if one aims at obtaining converged results that enable the separation between edge and bulk properties. Moreover, as a nanoribbon has two edges, electron transfer between the edges can occur in order to equilibrate the system, which complicates identifying the properties of single edges.   

In this paper we formulate a Green's function technique for calculating edge states of (semi-)infinite 2D systems with a single well-defined edge or grain boundary. We express bulk, edge and boundary Green's functions in terms of Bloch matrices. The latter are constructed from the solutions of a quadratic eigenvalue equation, which gives the traveling and evanescent Bloch states of the system. Electronic and structural reconstructions of edges or grain boundaries are easily incorporated in the technique. The formalism can be implemented in any localized basis set representation of the Hamiltonian.

Here we use it to calculate edge and grain boundary states of \MX\ monolayers by means of tight-binding models. A simple three-band model is employed to explore the electronic structures of the basic pristine zigzag and armchair edges in \MX. Within this model the zigzag edges are metallic in their charge neutral state, whereas the armchair edge is semiconducting. The three-band model is also used to analyze the electronic structures of \MS\ edges with a more general orientation between zigzag and armchair; these edges are generally metallic. The same model is applied to  obtain electronic structures of grain boundaries and of structurally modifed edges. 

The three-band model captures only part of \MX\ edge electronic structure. A more complete picture of the rich electronic structure of \MX\ edges  is obtained from an eleven-band tight-binding model comprising the $d$ metal valence orbitals and the $p$ valence orbitals of the chalcogen atoms. Edge states in the band gap with even symmetry (with respect to a mirror through the plane of M atoms) are qualitatively similar to those found in the three-band model. As in the latter model, charge neutral zigzag edges are metallic and the armchair edge is semiconducting in the eleven-band model. Unlike the three-band model, however, the charge neutrality level is fixed by edge states of odd symmetry in the band gap. The odd states have a character that reflects the dangling bonds on the edge atoms, unlike the even states, which have more of a mixed M-$d$ and X-$p$ character. The odd edge states are likely to be found near the Fermi level under experimental conditions, and can not be discarded when modeling \MX\ edges.

\acknowledgments{This work is part of the research program of the Foundation for Fundamental Research on Matter (FOM), which is part of the Netherlands Organization for Scientific Research (NWO).}

\appendix

\begin{widetext}
\section{Three-band tight-binding model}
\label{sec:3band}
Following Mattheis, the three-band tight binding model discussed in Sec~\ref{sec:tbmodels} only explicitly uses the sublattice of the M atoms and the $d$-orbitals of $A_1'$ and $E'$ symmetry, i.e., the orbitals that are symmetric with respect to $\sigma_h$, mirror symmetry in the \MS\ plane.\cite{Mattheis:prb73} We number these orbitals as 
\begin{equation}
d_{3z^2-r^2}= |0\rangle,\; d_{xy}= |1\rangle,\; d_{x^2-y^2}= |2\rangle.
\end{equation}
The matrix blocks $\mathbf{H}_{p,q}$ discussed in Sec.~\ref{sec:edges} are then $3\times 3$ matrices, where $\mathbf{H}_{p,q}$ denotes the real space Hamiltonian matrix block pertaining to the interaction between atoms in the unit cell at the origin and atoms in the unit cell at $p\mathbf{a}_1+q\mathbf{a}_2$. Using only nearest neighbor interactions, one has $|p|,|q|\leq 1$. Defining the tight-binding parameters $\epsilon_i=\langle i|h|i\rangle;\;i=0,1,2$ and $t_{ij}=\langle i|h|j\rangle;\;i,j=0,1,2$, and making use of the rotation properties of the $d$-orbitals, one derives the matrix blocks
\begin{equation}
\label{eq:a1}
\mathbf{H}_{0,0}=\left[\begin{array}{ccc}
\epsilon_0 & 0 & 0\\
0 & \epsilon_1 & 0 \\
0 & 0 & \epsilon_2 
\end{array}\right];\; \mathbf{H}_{1,0}= \left[\begin{array}{ccc}
t_{00} & \frac{\sqrt{3}}{2}t_{02}+\frac{1}{2}t_{01} & -\frac{1}{2}t_{02}+\frac{\sqrt{3}}{2}t_{01}\\
\frac{\sqrt{3}}{2}t_{02}-\frac{1}{2}t_{01} & \frac{1}{4}t_{11}+\frac{3}{4}t_{22} & \frac{\sqrt{3}}{4}(t_{11}-t_{22})-t_{12} \\
-\frac{1}{2}t_{02}-\frac{\sqrt{3}}{2}t_{01} & \frac{\sqrt{3}}{4}(t_{11}-t_{22})+t_{12} & \frac{1}{4}t_{22}+\frac{3}{4}t_{11} 
\end{array}\right];
\end{equation}
\begin{equation}
\label{eq:a3}
\mathbf{H}_{0,1}=\left[\begin{array}{ccc}
t_{00} & -t_{01} & t_{02} \\
t_{01} & t_{11} & -t_{12} \\
t_{02} & t_{12} & t_{22} 
\end{array}\right];\; \mathbf{H}_{1,1}= \left[\begin{array}{ccc}
t_{00} & -\frac{\sqrt{3}}{2}t_{02}-\frac{1}{2}t_{01} & -\frac{1}{2}t_{02}+\frac{\sqrt{3}}{2}t_{01}\\
-\frac{\sqrt{3}}{2}t_{02}+\frac{1}{2}t_{01} & \frac{1}{4}t_{11}+\frac{3}{4}t_{22} & -\frac{\sqrt{3}}{4}(t_{11}-t_{22})+t_{12} \\
-\frac{1}{2}t_{02}-\frac{\sqrt{3}}{2}t_{01} & -\frac{\sqrt{3}}{4}(t_{11}-t_{22})-t_{12} & \frac{1}{4}t_{22}+\frac{3}{4}t_{11} 
\end{array}\right].
\end{equation}

The hopping associated with matrix elements $t_{00},t_{02},t_{22},t_{11}$ is even with respect to inversion
and the one associated with $t_{01},t_{12}$ is odd. It is then easy to see that
\begin{equation}
\label{eq:a6}
\mathbf{H}_{-p,-q}=(\mathbf{H}_{p,q})^T
\end{equation}
The values of the tight-binding parameters are taken from Liu \textit{et al.},\cite{Liu:prb13} which have been ontained by fitting the tight-binding band structure of a \MX\ sheet to a DFT GGA/PBE band structure,\cite{Perdew:prl96} using the highest valence band and the two lowest conduction bands in the fit. GGA/PBE gives an optimized lattice parameter of $3.19$ \AA\ for \MS, which is 1-2 \% larger than the reported experimental values.\cite{Young:jpd68,Alhili:jcg15,Boker:prb64}  With this lattice parameter the calculated band gap is 1.63 eV, which is smaller than the experimental optical band gap of 1.85 eV.\cite{Mak:prl10} 

\section{Eleven-band tight-binding model}
\label{sec:11band}
The eleven-band tight-binding model of \MX\ uses a basis set composed of all five $d$-orbitals of the M atom, and the six $p$ orbitals of the two X atoms. Following the approach of Cappelluti \textit{et al.},\cite{Cappel:prb88,Gomez:nl13} we include next nearest neighbor interactions, and use the Slater-Koster two-center approximation for the hopping matrix elements.\cite{Slater:pr54} The real space matrix blocks $\mathbf{H}_{p,q}$ discussed in Sec.~\ref{sec:edges} then have the form 
\begin{equation}
\label{eq:a8}
\mathbf{H}_{0,0}=\left[\begin{array}{cc}
\boldsymbol{\epsilon}_d & \mathbf{t}_{dp}^{0,0} \\
(\mathbf{t}_{dp}^{0,0})^T & \boldsymbol{\epsilon}_p \\ 
\end{array}\right],\; \mathbf{H}_{1,0}=\left[\begin{array}{cc}
\mathbf{t}_{dd}^{1,0} & \mathbf{0} \\
\mathbf{0} & \mathbf{t}_{pp}^{1,0} \\ 
\end{array}\right],\; \mathbf{H}_{0,1}=\left[\begin{array}{cc}
\mathbf{t}_{dd}^{0,1} & \mathbf{0} \\
(\mathbf{t}_{dp}^{0,1})^T & \mathbf{t}_{pp}^{0,1} \\ 
\end{array}\right],\; \mathbf{H}_{1,1}=\left[\begin{array}{cc}
\mathbf{t}_{dd}^{1,1} & \mathbf{0} \\
(\mathbf{t}_{dp}^{1,1})^T & \mathbf{t}_{pp}^{1,1} \\ 
\end{array}\right],
\end{equation}
with the remaining blocks constructed according to Eq.~\ref{eq:a6}. The bulk Hamiltonian for an infinite layer with two-dimensional translation symmetry is written in this notation as
\begin{equation}
\mathbf{H}(k_1,k_2) = \mathbf{H}_{0,0} + \mathbf{A} + \mathbf{A}^\dagger;\;\; \mathbf{A} = \mathbf{H}_{1,0} e^{i2\pi k_1} + \mathbf{H}_{0,1} e^{i2\pi k_2} + \mathbf{H}_{1,1} e^{i2\pi (k_1 + k_2)}.
\end{equation}
The parameters in the model can be found by fitting the tight-binding band structure obtained with this bulk Hamiltonian to a band structure obtained from a DFT calculation. The model turns out to be too restrictive to obtain a satisfactory fit for all eleven bands with one parameter set. A good fit can however be obtained if we divide the bands into a set of even symmetry, and a set of odd symmetry, and use different parameters for the two sets. Mirror symmetry $\sigma_h$ in the \MX\ plane holds for monolayers, as well as for edges, so states that are even or odd with respect to $\sigma_h$, can be treated separately. In Sec.~\ref{sec:evenmatrix} we give the expressions for the matrix blocks $\boldsymbol{\epsilon}_a$ and $\mathbf{t}_{ab}^{p,q}$ in Eq.~\ref{eq:a8} for the even states, and in Sec.~\ref{sec:oddmatrix} for the odd states.
  
\subsection{Even states}
\label{sec:evenmatrix}
The even states are composed of orbitals of $E'$ and $A_1'$ symmetry, 
\begin{eqnarray}
&&d_{3z^2-r^2}= |0\rangle,\; d_{xy}= |1\rangle,\; d_{x^2-y^2}= |2\rangle, \nonumber \\
&&\frac{1}{\sqrt{2}}\left[p_x(\mathrm{X}_1)+p_x(\mathrm{X}_2)\right]= |3\rangle,\; \frac{1}{\sqrt{2}}\left[p_y(\mathrm{X}_1)+p_y(\mathrm{X}_2)\right]= |4\rangle,\; \frac{1}{\sqrt{2}}\left[p_z(\mathrm{X}_1)-p_z(\mathrm{X}_2)\right]=|5\rangle.
\end{eqnarray}
The matrix blocks $\boldsymbol{\epsilon}_a$ and $\mathbf{t}_{ab}^{p,q}$ in Eq.~\ref{eq:a8} are then all $3\times 3$
\begin{equation}
\label{eq:a9}
\boldsymbol{\epsilon}_d=\left[\begin{array}{ccc}
\epsilon_{3z^2-r^2} & 0 & 0 \\
0 & \epsilon_{xy} & 0 \\
0 & 0 & \epsilon_{xy} 
\end{array}\right],\; \;  \boldsymbol{\epsilon}_p=\left[\begin{array}{ccc}
\epsilon_{p_x}+V_{pp\pi} & 0 & 0 \\
0 & \epsilon_{p_x}+V_{pp\pi} & 0 \\
0 & 0 & \epsilon_{p_z}-V_{pp\sigma} 
\end{array}\right],
\end{equation}
where $\epsilon_a$ are the on-site orbital energies and $V_{ab\alpha}$ are the Slater-Koster two-center integrals.\cite{Slater:pr54}
Similarly,
\begin{equation}
\label{eq:a10}
\mathbf{t}_{dp}^{0,0}=\frac{\sqrt{2}}{7\sqrt{7}}\left[\begin{array}{ccc}
9V_{pd\pi}-\sqrt{3}V_{pd\sigma} & 3\sqrt{3}V_{pd\pi}-V_{pd\sigma} & 12V_{pd\pi}+\sqrt{3}V_{pd\sigma} \\
-V_{pd\pi}-3\sqrt{3}V_{pd\sigma} & -5\sqrt{3}V_{pd\pi}-3V_{pd\sigma} & -6V_{pd\pi}+3\sqrt{3}V_{pd\sigma} \\
-5\sqrt{3}V_{pd\pi}-3V_{pd\sigma} & 9V_{pd\pi}-\sqrt{3}V_{pd\sigma} & -2\sqrt{3}V_{pd\pi}+3V_{pd\sigma} 
\end{array}\right], 
\end{equation}
\begin{equation}
\label{eq:a12}
\mathbf{t}_{dd}^{1,0}=\left[\begin{array}{ccc}
\frac{1}{4}(3V_{dd\delta}+V_{dd\sigma}) & \frac{3}{8}(-V_{dd\delta}+V_{dd\sigma}) & \frac{\sqrt{3}}{8}(-V_{dd\delta}+V_{dd\sigma}) \\
\frac{3}{8}(-V_{dd\delta}+V_{dd\sigma}) & \frac{1}{16}(3V_{dd\delta}+4V_{pp\pi}+9V_{dd\sigma}) & \frac{\sqrt{3}}{16}(V_{dd\delta}-4V_{dd\pi}+3V_{dd\sigma}) \\
\frac{\sqrt{3}}{8}(-V_{dd\delta}+V_{dd\sigma}) & \frac{\sqrt{3}}{16}(V_{dd\delta}-4V_{dd\pi}+3V_{dd\sigma}) & \frac{1}{16}(V_{dd\delta}+12V_{pp\pi}+3V_{dd\sigma}) 
\end{array}\right], 
\end{equation}
\begin{equation}
\label{eq:a13}
\mathbf{t}_{pp}^{1,0}=\left[\begin{array}{ccc}
\frac{1}{4}(3V_{pp\pi}+V_{pp\sigma}) & \frac{\sqrt{3}}{4}(V_{pp\pi}-V_{pp\sigma}) & 0 \\
\frac{\sqrt{3}}{4}(V_{pp\pi}-V_{pp\sigma}) & \frac{1}{4}(V_{pp\pi}+3V_{pp\sigma}) & 0 \\
0 & 0 & V_{pp\pi} 
\end{array}\right], 
\end{equation}
\begin{equation}
\label{eq:a15}
\mathbf{t}_{dd}^{0,1}=\left[\begin{array}{ccc}
\frac{1}{4}(3V_{dd\delta}+V_{dd\sigma}) & \frac{3}{8}(V_{dd\delta}-V_{dd\sigma}) & \frac{\sqrt{3}}{8}(-V_{dd\delta}+V_{dd\sigma}) \\
\frac{3}{8}(V_{dd\delta}-V_{dd\sigma}) & \frac{1}{16}(3V_{dd\delta}+4V_{pp\pi}+9V_{dd\sigma}) & \frac{\sqrt{3}}{16}(-V_{dd\delta}+4V_{dd\pi}-3V_{dd\sigma}) \\
\frac{\sqrt{3}}{8}(-V_{dd\delta}+V_{dd\sigma}) & \frac{\sqrt{3}}{16}(-V_{dd\delta}+4V_{dd\pi}-3V_{dd\sigma}) & \frac{1}{16}(V_{dd\delta}+12V_{pp\pi}+3V_{dd\sigma}) 
\end{array}\right], 
\end{equation}
\begin{equation}
\label{eq:a16}
\mathbf{t}_{pp}^{0,1}=\left[\begin{array}{ccc}
\frac{1}{4}(3V_{pp\pi}+V_{pp\sigma}) & \frac{\sqrt{3}}{4}(-V_{pp\pi}+V_{pp\sigma}) & 0 \\
\frac{\sqrt{3}}{4}(-V_{pp\pi}+V_{pp\sigma}) & \frac{1}{4}(V_{pp\pi}+3V_{pp\sigma}) & 0 \\
0 & 0 & V_{pp\pi} 
\end{array}\right],\;\; 
\mathbf{t}_{dp}^{0,1}= \frac{2\sqrt{2}}{7\sqrt{7}}\left[\begin{array}{ccc}
0 & -3\sqrt{3}V_{pd\pi}+V_{pd\sigma} & 6V_{pd\pi}+\frac{1}{2}\sqrt{3}V_{pd\sigma} \\
V_{pd\pi} & 0 & 0  \\
0 & -3V_{pd\pi}-2\sqrt{3}V_{pd\sigma} & 2\sqrt{3}V_{pd\pi}-3V_{pd\sigma} 
\end{array}\right], 
\end{equation}
\begin{equation}
\label{eq:a19}
\mathbf{t}_{dd}^{1,1}=\left[\begin{array}{ccc}
\frac{1}{4}(3V_{dd\delta}+V_{dd\sigma}) & 0 & \frac{\sqrt{3}}{4}(V_{dd\delta}-V_{dd\sigma}) \\
0 & V_{dd\pi} & 0 \\
\frac{\sqrt{3}}{4}(V_{dd\delta}-V_{dd\sigma}) & 0 & \frac{1}{4}(V_{dd\delta}+3V_{dd\sigma}) 
\end{array}\right], \;\;
\mathbf{t}_{pp}^{1,1}=\left[\begin{array}{ccc}
V_{pp\sigma} & 0 & 0 \\
0 & V_{pp\pi} & 0 \\
0 & 0 & V_{pp\pi} 
\end{array}\right], 
\end{equation}
\begin{equation}
\label{eq:a21}
\mathbf{t}_{dp}^{1,1}=\frac{\sqrt{2}}{7\sqrt{7}}\left[\begin{array}{ccc}
-9V_{pd\pi}+\sqrt{3}V_{pd\sigma} & 3\sqrt{3}V_{pd\pi}-V_{pd\sigma} & 12V_{pd\pi}+\sqrt{3}V_{pd\sigma} \\
-V_{pd\pi}-3\sqrt{3}V_{pd\sigma} & 5\sqrt{3}V_{pd\pi}+3V_{pd\sigma} & 6V_{pd\pi}-3\sqrt{3}V_{pd\sigma} \\
5\sqrt{3}V_{pd\pi}+3V_{pd\sigma} & 9V_{pd\pi}-\sqrt{3}V_{pd\sigma} & -2\sqrt{3}V_{pd\pi}+3V_{pd\sigma} 
\end{array}\right].
\end{equation}
We use values of the parameters as obtained by Rostami \textit{at al.} from fitting the even tight-binding bands to DFT GGA/PBE bands.\cite{Rostami:prb15} A lattice parameter of $3.16$ \AA\ for \MS\ has been used in these calculations, which is 1\% smaller than the optimized GGA/PBE value, but close to the reported experimental values. It results in a calculated band gap of 1.76 eV, which is slightly smaller than the experimental optical band gap of 1.85 eV.\cite{Mak:prl10} We have shifted the tight-binding bands such, that the zero of energy coincides with the top of the valence band.

\subsection{Odd states}
\label{sec:oddmatrix}
The odd states are composed of orbitals of $E''$ and $A_1''$ symmetry, 
\begin{eqnarray}
&&d_{xz}= |1'\rangle,\; d_{yz}= |2'\rangle, \nonumber \\
&&\frac{1}{\sqrt{2}}\left[p_x(\mathrm{X}_1)-p_x(\mathrm{X}_2)\right]= |3'\rangle,\; \frac{1}{\sqrt{2}}\left[p_y(\mathrm{X}_1)-p_y(\mathrm{X}_2)\right]= |4'\rangle,\; \frac{1}{\sqrt{2}}\left[p_z(\mathrm{X}_1)+p_z(\mathrm{X}_2)\right]=|5'\rangle.
\end{eqnarray}
The matrix blocks $\boldsymbol{\epsilon}_a$ in Eq.~\ref{eq:a8} are $2\times 2$ if $a=d$ and $3\times 3$ if $a=p$, whereas the matrix blocks $\mathbf{t}_{ab}^{p,q}$ are are $2\times 2$ if $ab=dd$, $3\times 3$ if $ab=pp$, and $2\times 3$ if $ab=dp$ 
\begin{equation}
\label{eq:a24}
\boldsymbol{\epsilon}_{d}=\left[\begin{array}{cc}
\epsilon'_{xz} & 0 \\
0 & \epsilon'_{xz} 
\end{array}\right],\; \;  \boldsymbol{\epsilon}_{p}=\left[\begin{array}{ccc}
\epsilon'_{p_x}-V'_{pp\pi} & 0 & 0 \\
0 & \epsilon'_{p_x}-V'_{pp\pi} & 0 \\
0 & 0 & \epsilon'_{p_z}+V'_{pp\sigma} 
\end{array}\right],
\end{equation}
\begin{equation}
\label{eq:a25}
\mathbf{t}_{dp}^{0,0}=\frac{\sqrt{2}}{7\sqrt{7}}\left[\begin{array}{ccc}
\sqrt{3}V'_{pd\pi}+9V'_{pd\sigma} & -6V'_{pd\pi}+3\sqrt{3}V'_{pd\sigma} & -\sqrt{3}V'_{pd\pi}-9V'_{pd\sigma} \\
-6V'_{pd\pi}+3\sqrt{3}V'_{pd\sigma} & 5\sqrt{3}V'_{pd\pi}+3V'_{pd\sigma} & -V'_{pd\pi}-3\sqrt{3}V'_{pd\sigma}
\end{array}\right], 
\end{equation}
\begin{equation}
\label{eq:a26}
\mathbf{t}_{dd}^{1,0}=\left[\begin{array}{cc}
\frac{1}{4}(3V'_{dd\delta}+V'_{dd\pi}) & \frac{\sqrt{3}}{4}(V'_{dd\delta}-V'_{dd\pi}) \\
\frac{\sqrt{3}}{4}(V'_{dd\delta}-V'_{dd\pi}) & \frac{1}{4}(V'_{dd\delta}+3V'_{pp\pi)} 
\end{array}\right],\; 
\mathbf{t}_{pp}^{1,0}=\left[\begin{array}{ccc}
\frac{1}{4}(3V'_{pp\pi}+V'_{pp\sigma}) & \frac{\sqrt{3}}{4}(V'_{pp\pi}-V'_{pp\sigma}) & 0 \\
\frac{\sqrt{3}}{4}(V'_{pp\pi}-V'_{pp\sigma}) & \frac{1}{4}(V'_{pp\pi}+3V'_{pp\sigma}) & 0 \\
0 & 0 & V'_{pp\pi} 
\end{array}\right], 
\end{equation}
\begin{equation}
\label{eq:a27}
\mathbf{t}_{dd}^{0,1}=\left[\begin{array}{cc}
\frac{1}{4}(3V'_{dd\delta}+V'_{dd\pi}) & \frac{\sqrt{3}}{4}(-V'_{dd\delta}+V'_{dd\pi}) \\
\frac{\sqrt{3}}{4}(-V'_{dd\delta}+V'_{dd\pi}) & \frac{1}{4}(V'_{dd\delta}+3V'_{dd\pi}) 
\end{array}\right],\; \mathbf{t}_{pp}^{0,1}=\left[\begin{array}{ccc}
\frac{1}{4}(3V'_{pp\pi}+V'_{pp\sigma}) & \frac{\sqrt{3}}{4}(-V'_{pp\pi}+V'_{pp\sigma}) & 0 \\
\frac{\sqrt{3}}{4}(-V'_{pp\pi}+V'_{pp\sigma}) & \frac{1}{4}(V'_{pp\pi}+3V'_{pp\sigma}) & 0 \\
0 & 0 & V'_{pp\pi} 
\end{array}\right], 
\end{equation}
\begin{equation}
\label{eq:a28}
\mathbf{t}_{dp}^{0,1}=\frac{\sqrt{2}}{7\sqrt{7}}\left[\begin{array}{ccc}
\sqrt{3}V'_{pd\pi} & 0 & 0  \\
0 & -\sqrt{3}V'_{pd\pi}+12V'_{pd\sigma} & 2V'_{pd\pi}+6\sqrt{3}V'_{pd\sigma}
\end{array}\right], 
\end{equation}
\begin{equation}
\label{eq:a29}
\mathbf{t}_{dd}^{1,1}=\left[\begin{array}{cc}
V'_{dd\pi} & 0  \\
0 & V'_{dd\delta} 
\end{array}\right], \;\;
\mathbf{t}_{pp}^{1,1}=\left[\begin{array}{ccc}
V'_{pp\sigma} & 0 & 0 \\
0 & V'_{pp\pi} & 0 \\
0 & 0 & V'_{pp\pi} 
\end{array}\right], 
\end{equation}
\begin{equation}
\label{eq:a30}
\mathbf{t}_{dp}^{1,1}=\frac{\sqrt{2}}{7\sqrt{7}}\left[\begin{array}{ccc}
\sqrt{3}V'_{pd\pi}+9V'_{pd\sigma} & 6V'_{pd\pi}-3\sqrt{3}V'_{pd\sigma} & \sqrt{3}V'_{pd\pi}+9V'_{pd\sigma} \\
6V'_{pd\pi}-3\sqrt{3}V'_{pd\sigma} & 5\sqrt{3}V'_{pd\pi}+3V'_{pd\sigma} & -V'_{pd\pi}-3\sqrt{3}V'_{pd\sigma}
\end{array}\right].
\end{equation}
We obtain values of the parameters by fitting the tight-binding bands of odd symmetry to bands obtained from a density functional theory (DFT) calculation with the generalized gradient GGA/PBE functional,\cite{Perdew:prl96,Kresse:prb96,Kresse:prb99} using the same lattice parameter as for the even bands. The optimal parameters are given in table~\ref{tab:oddparam}, and the quality of the fit can be judged from Fig.~\ref{fig:oddbulk}.
\end{widetext}

\begin{table}[h]
\caption{Values of the tight-binding parameters (eV) used for the band of odd symmetry.}
\begin{ruledtabular}
\begin{tabular}{ldld}
$\epsilon'_{p_x}$   & $-2.188$ & $\epsilon'_{p_z}$  & $-0.682$  \\    
$\epsilon'_{xz}$  & $0.604$  & & \\
$V'_{dd\pi}$ & $-0.075$ & $V'_{dd\delta}$ & $0.051$  \\
$V'_{pp\sigma}$ & $1.166$ & $V'_{pp\pi}$ & $-0.389$  \\ 
$V'_{pd\sigma}$ & $2.115$ & $V'_{pd\pi}$ & $-0.626$            
\end{tabular}
\end{ruledtabular} 
\label{tab:oddparam}
\end{table}

\begin{figure}[h]
\includegraphics[width=0.9\columnwidth]{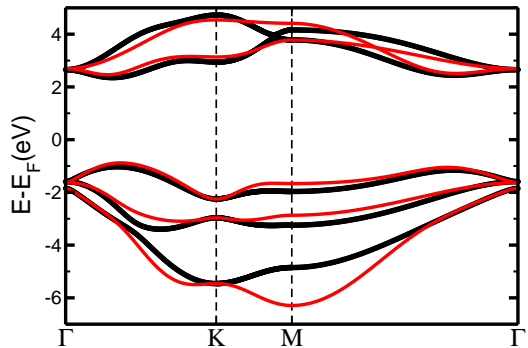}
\caption{(Color online) GGA/PBE band structure (black) of the odd symmetry bands of \MS; (red) tight-binding fit.} 
 \label{fig:oddbulk}
\end{figure}  

%

\end{document}